\documentclass[final,5p,times,twocolumn]{elsarticle}
\usepackage{amsmath,amsopn,amscd,amsthm,amssymb,xypic,rotating}
\newcommand{\bp}            {}

\newcommand{\separate}      {\bigskip}
\newcommand{\double}       {}
\newcommand{\single}       {}
\newcommand{\dqt}[1]        {``{#1}''}
\def\shortciteA{\cite}
\def\shortcite{\cite}
\def\citeA{\cite}
\def\figwidth{0.8\columnwidth}

\newcommand{\tset}[1]      {\{{#1}\}}                     
\newcommand{\tbag}[1]      {\{\!\!|{#1}\}\!\!\!|}         
\newcommand{\tlist}[1]     {[{#1}]}                       
\newcommand{\ttree}[1]     {{\frak t}({#1})}              
\newcommand{\cmd}[1]      {\underline{{#1}}}
\newcommand{\cb}          {\begin{tabbing}MMMMM\=MM\=MM\=MM\=MM\=MM\=MM\=MM\=MM\=MM\= \kill}
\newcommand{\ce}          {\end{tabbing}}

\newcommand{\HInsert}[2]   {\centerline {\immediate\pdfximage width #2  {#1}\pdfrefximage\pdflastximage}}
\newcommand{\VInsert}[2]   {\centerline {\immediate\pdfximage height #2 {#1}\pdfrefximage\pdflastximage}}

\newcommand{\boxx}[1]       {\left[\!\left[{#1}\right]\!\right]}

\journal{ArXiv}

\begin{document}

\begin{frontmatter}

\title{Opinion formation in a locally interacting community with recommender}
\author{Simone Santini}
\address{Universidad Aut\'onoma de Madrid, Madrid, Spain}
\ead{simone.santini@uam.es}

\begin{abstract}
We present a user of model interaction based on the physics of kinetic
exchange, and extend it to individuals placed in a grid with local
interaction. We show with numerical analysis and partial analytical
results that the critical symmetry breaking transitions and
percolation effects typical of the full interaction model do not take
place if the range of interaction is limited, allowing for the
co-existence of majorty and minority opinions in the same community.

We then introduce a 
peer recommender system in the model, showing that,
even with very local iteraction and a small probability of appeal to
the recommender, its presence is sufficient to make both symmetry
breaking and percolation reappear. This seems to indicate that one
effect of a recommendation system is to uniform the opinions of a
community, reducing minority opinions or making them
disappear. Although the recommender system does uniform the community
opinion, it doesn't constrain it, in the sense that all opinions have
the same probability of becoming the dominating one. We do a partial
study, however, that suggests that a \dqt{mischievous} recommender might
be able to bias a community so that one opinion will emerge over the
opposite with overwhelming probability.
\end{abstract}


\end{frontmatter}
\begin{keyword}
Kinetic exchange \sep opinion formation model \sep recommender systems
\end{keyword}



\section{Introduction}
Standard models of opinion formation have generally been from economic
models based on game theory \cite{dimare:06}: free agents interact
with each other interchanging \emph{opinions} in lieu of wealth. The
main result of this work is that opinion exchange systems present a
Nash equilibrium \cite{nash:51} in which each agent holds the
\dqt{best} opinion possible (that is, the one that leads to more
fruitful engagement) given the opinions held by the rest of the group.

More recently, an alternative model has emerged in the form of
\emph{kinetic exchange models} \cite{toscani:06,lallouache:10} in
which the agents are considered as free wandering particles (viz. one
assumes full interaction) that, upon meeting, influence each other,
exchanging, in part, their opinions. The advantage of this model is
that it is formally very similar to gas dynamic models
\cite{cercignani:94} to obtain analytical results on the behavior of
the model.

These models assume no external influence on the opinion of the
individuals: the opinion of an individual changes only through a
process in which individuals reach a fair compromise after exchanging
opinions \cite{bennaim:03,bennaim:03b}. Their most interesting
characteristic is a symmetry breaking transition for a specific value
of a \emph{conviction} parameter, at which point the generally neutral
opinion is transformed into a strongly polarized unanimous opinion
\cite{lallouache:10}. For low values of the conviction parameters,
several clusters of different opinions can coexist in the comunity,
but around the critical value the main cluster \emph{percolates} into
the whole community, unifying the opinion.

In this paper, we extend this model in two ways. First, we place the
individuals in a rectangular grid, and allow only interactions at a
limited range. We show that with this change, the percolation of the
main cluster disappear if the range is less than about half the linear
size of the community.  The symmetry breaking transition doesn't
disappear, but it becomes less and less polarized, allowing the
coexistence of dissenting opinions even though the average opinion is
not neutral. That is, limited interactions allow the coexistence of a
majority opinion and a minority one. (Such a phenomenon has been
observed in other types of models, such as \cite{kacperski:99}.) We
then show that the presence of the recommender system is sufficient to
restore uniformity of opinion even with very limited interaction
range. The symmetry breaking transition and the percolation of the
main cluster take place as in the fully connected system even for very
limited interaction ranges and for very limited interactions with the
recommender.

Finally, we study the consequence of the presence of a
\emph{mischievous} recommender system, one that tries to steer opinion
towards one of the extremes.

\section{The basic model}
Assume a set of $N$ individuals, each one having, at time $t$, an
opinion $O_i(t)\in[-1,1]$ ($i=1,\ldots,N$). When individuals $i$ and
$j$ interact, the opinion of each one changes as a consequence of a
negotiation process that tends to make them more similar to one
another. If $\lambda_i\in[0,1]$ is the \emph{conviction} of individual
$i$, that is, the strength with which an individual holds her opinions
and if we define the boxing operator
\begin{equation}
\boxx{x} = 
\begin{cases}
-1 & \mbox{if $x<-1$} \\
x & \mbox{if $-1\le x \le 1$} \\
1 & \mbox{if $x>1$} 
\end{cases}
\end{equation}
then we can model the interchange between $i$ and $j$
\cite{toscani:06} as:
\begin{equation}
\label{basicint}
\begin{aligned}
O_i(t+1) &= \boxx{\lambda_i O_i + \lambda_j \epsilon_t O_j(t)} \\
O_j(t+1) &= \boxx{\lambda_j O_j + \lambda_i \epsilon_t^\prime O_i(t)}
\end{aligned}
\end{equation}
where $\epsilon_t$ and $\epsilon_t^\prime$ are annealed (time-varying)
variables: uncorrelated stochastic processes with uniform distribution
in $[0,1]$. Here we are assuming that the conviction of an individual
is equal to his proselytizing power, viz. to her capacity to convince
others. There is a more complex model in which different parameters
are introduced for conviction ($\lambda_i$) and proselytism ($\mu_i$)
but whose behavior is similar to the simpler one. If we assume
uniformity of conviction (viz. a community without a strong leader)
then the one-side interaction can be written as:
\begin{equation}
\label{model1}
O_i(t+1) = \boxx{\lambda (O_i + \epsilon_t O_j(t))} 
\end{equation}
The basic model is assumed to be fully connected, that is, any agent
can interact with any other agent. This system is described by an
order parameter, which is simply the average opinion among all
individuals: 
\begin{equation}
\label{order}
O(t) = \frac{1}{N} \sum_{i=1}^N O_i(t)
\end{equation}
The analytical study of the model (\ref{model1}) is in general
impossible, although mean-field solutions can be found in special
cases \cite{biswas:11}. In the hypothesis of full interaction, the
time evolution of the order parameter is described adequately by the
equation
\begin{equation}
\label{evol}
O(t+1) = \boxx{\lambda(1+\epsilon O(t))}
\end{equation}
We can study the stochastic map (\ref{evol}) in terms of random
walks. Defining $S(t)=\log\,|O(t)|$, eq. (\ref{evol}) can be written as
\begin{equation}
S(t+1) = S(t) + \nu
\end{equation}
where $\nu=\log\,\lambda(1+\epsilon)$. The presence of the boxing
function entails that we are actually describing a random walk with a
reflecting boundaty at $S=0$ (when $S=0$ the box function will make it
bounce back into negative values). Depending on the value $\lambda$
the random walk can be biased towards positive or towards negative
values, and there is a critical value $\lambda_c$ at which it is
unbiased. Averaging independently over the two terms of the sum, onec
can estimate the critical point \cite{toscani:06,chowdhurry:11}. The
walk is unbiased if $\langle\lambda\rangle=0$, viz.
\begin{equation}
\int_0^1 \log\,\lambda_c(1+\epsilon) d\epsilon = 0
\end{equation}
giving $\lambda_c=e/4$. We can estimate the dependence of $O_a$ on
$\lambda$ (at steady state, as an ensemble average over all
trajectories) by first estimating the average \dqt{return time} $T$,
that is, the time between two consecutive bounces at the reflecting
boundary. Since $\epsilon$ is uniformly distributed, after a bounce
the walker will go on a verage to a position $(\lambda+1)/2$. The
average contribution of each step of the walk is given by
$\int_0^1\log[\lambda(1+\epsilon)]d\epsilon=\log(\lambda/\lambda_c)$. This
is a measure of the bias of the walk which, for
$\lambda\rightarrow\lambda_c$ varies as $(\lambda-\lambda_c)$. $T$ of
these steps will take us back (on average) to the boundary, so
multiplying this value by itself $T$ time we should go back from the
bounce position to 1, that is:
\begin{equation}
\frac{\lambda+1}{2} \left(\frac{\lambda}{\lambda_c}\right)^T = 1
\end{equation}
which yields
\begin{equation}
T = - \frac{\log\,\lambda}{\log\,\lambda-\log\,\lambda_c}
  \approx - \frac{\log\,\lambda}{\lambda-\lambda_c}
\end{equation}
where the approximation is valid for
$\lambda\rightarrow\lambda_c$. The steady state average of $S$ is
expected to be
\begin{equation}
S_a \sim \sqrt{T}\log\,\lambda
\end{equation}
that is
\begin{equation}
S_a = k\sqrt{T}\log\,\lambda
\end{equation}
where $k$ is a constant to the determined by fitting the data. This
gives an approximation
\begin{equation}
\label{prediction}
|O_a| = \exp(-k|\log\,\lambda|^{\frac{3}{2}} (\lambda-\lambda_c)^{\frac{1}{2}})
\end{equation}
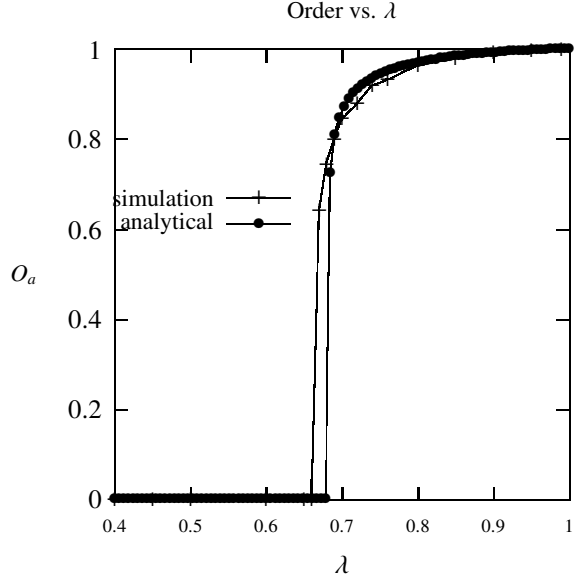
\begin{figure}[tb]
  \setlength{\unitlength}{0.240900pt}
  \ifx\plotpoint\undefined\newsavebox{\plotpoint}\fi
  \sbox{\plotpoint}{\rule[-0.200pt]{0.400pt}{0.400pt}}%
  \begin{picture}(960,960)(0,0)
    \sbox{\plotpoint}{\rule[-0.200pt]{0.400pt}{0.400pt}}%
    \put(182.0,131.0){\rule[-0.200pt]{4.818pt}{0.400pt}}
    \put(162,131){\makebox(0,0)[r]{ 0}}
    \put(868.0,131.0){\rule[-0.200pt]{4.818pt}{0.400pt}}
    \put(182.0,272.0){\rule[-0.200pt]{4.818pt}{0.400pt}}
    \put(162,272){\makebox(0,0)[r]{ 0.2}}
    \put(868.0,272.0){\rule[-0.200pt]{4.818pt}{0.400pt}}
    \put(182.0,413.0){\rule[-0.200pt]{4.818pt}{0.400pt}}
    \put(162,413){\makebox(0,0)[r]{ 0.4}}
    \put(868.0,413.0){\rule[-0.200pt]{4.818pt}{0.400pt}}
    \put(182.0,554.0){\rule[-0.200pt]{4.818pt}{0.400pt}}
    \put(162,554){\makebox(0,0)[r]{ 0.6}}
    \put(868.0,554.0){\rule[-0.200pt]{4.818pt}{0.400pt}}
    \put(182.0,695.0){\rule[-0.200pt]{4.818pt}{0.400pt}}
    \put(162,695){\makebox(0,0)[r]{ 0.8}}
    \put(868.0,695.0){\rule[-0.200pt]{4.818pt}{0.400pt}}
    \put(182.0,836.0){\rule[-0.200pt]{4.818pt}{0.400pt}}
    \put(162,836){\makebox(0,0)[r]{ 1}}
    \put(868.0,836.0){\rule[-0.200pt]{4.818pt}{0.400pt}}
    \put(182.0,131.0){\rule[-0.200pt]{0.400pt}{4.818pt}}
    \put(182,90){\makebox(0,0){\scriptsize 0.4}}
    \put(182.0,816.0){\rule[-0.200pt]{0.400pt}{4.818pt}}
    \put(300.0,131.0){\rule[-0.200pt]{0.400pt}{4.818pt}}
    \put(300,90){\makebox(0,0){\scriptsize 0.5}}
    \put(300.0,816.0){\rule[-0.200pt]{0.400pt}{4.818pt}}
    \put(417.0,131.0){\rule[-0.200pt]{0.400pt}{4.818pt}}
    \put(417,90){\makebox(0,0){\scriptsize 0.6}}
    \put(417.0,816.0){\rule[-0.200pt]{0.400pt}{4.818pt}}
    \put(535.0,131.0){\rule[-0.200pt]{0.400pt}{4.818pt}}
    \put(535,90){\makebox(0,0){\scriptsize 0.7}}
    \put(535.0,816.0){\rule[-0.200pt]{0.400pt}{4.818pt}}
    \put(653.0,131.0){\rule[-0.200pt]{0.400pt}{4.818pt}}
    \put(653,90){\makebox(0,0){\scriptsize 0.8}}
    \put(653.0,816.0){\rule[-0.200pt]{0.400pt}{4.818pt}}
    \put(770.0,131.0){\rule[-0.200pt]{0.400pt}{4.818pt}}
    \put(770,90){\makebox(0,0){\scriptsize 0.9}}
    \put(770.0,816.0){\rule[-0.200pt]{0.400pt}{4.818pt}}
    \put(888.0,131.0){\rule[-0.200pt]{0.400pt}{4.818pt}}
    \put(888,90){\makebox(0,0){\scriptsize 1}}
    \put(888.0,816.0){\rule[-0.200pt]{0.400pt}{4.818pt}}
    \put(182.0,131.0){\rule[-0.200pt]{0.400pt}{169.834pt}}
    \put(182.0,131.0){\rule[-0.200pt]{170.075pt}{0.400pt}}
    \put(888.0,131.0){\rule[-0.200pt]{0.400pt}{169.834pt}}
    \put(182.0,836.0){\rule[-0.200pt]{170.075pt}{0.400pt}}
    \put(41,483){\makebox(0,0){\small{$O_a$}}}
    \put(535,29){\makebox(0,0){$\lambda$}}
    \put(535,898){\makebox(0,0){\small{Order vs. $\lambda$}}}
    \put(336,604){\makebox(0,0)[r]{\small{simulation}}}
    \put(356.0,604.0){\rule[-0.200pt]{24.090pt}{0.400pt}}
    \put(182,131){\usebox{\plotpoint}}
    \multiput(488.58,131.00)(0.492,19.538){21}{\rule{0.119pt}{15.233pt}}
    \multiput(487.17,131.00)(12.000,422.382){2}{\rule{0.400pt}{7.617pt}}
    \multiput(500.58,585.00)(0.492,3.373){19}{\rule{0.118pt}{2.718pt}}
    \multiput(499.17,585.00)(11.000,66.358){2}{\rule{0.400pt}{1.359pt}}
    \multiput(511.58,657.00)(0.492,1.616){21}{\rule{0.119pt}{1.367pt}}
    \multiput(510.17,657.00)(12.000,35.163){2}{\rule{0.400pt}{0.683pt}}
    \multiput(523.58,695.00)(0.492,1.401){21}{\rule{0.119pt}{1.200pt}}
    \multiput(522.17,695.00)(12.000,30.509){2}{\rule{0.400pt}{0.600pt}}
    \multiput(535.00,728.58)(0.498,0.496){45}{\rule{0.500pt}{0.120pt}}
    \multiput(535.00,727.17)(22.962,24.000){2}{\rule{0.250pt}{0.400pt}}
    \multiput(559.58,752.00)(0.496,0.609){43}{\rule{0.120pt}{0.587pt}}
    \multiput(558.17,752.00)(23.000,26.782){2}{\rule{0.400pt}{0.293pt}}
    \multiput(582.00,780.59)(1.368,0.489){15}{\rule{1.167pt}{0.118pt}}
    \multiput(582.00,779.17)(21.579,9.000){2}{\rule{0.583pt}{0.400pt}}
    \multiput(606.00,789.58)(1.028,0.496){43}{\rule{0.917pt}{0.120pt}}
    \multiput(606.00,788.17)(45.096,23.000){2}{\rule{0.459pt}{0.400pt}}
    \multiput(653.00,812.58)(2.713,0.492){19}{\rule{2.209pt}{0.118pt}}
    \multiput(653.00,811.17)(53.415,11.000){2}{\rule{1.105pt}{0.400pt}}
    \multiput(711.00,823.59)(3.407,0.489){15}{\rule{2.722pt}{0.118pt}}
    \multiput(711.00,822.17)(53.350,9.000){2}{\rule{1.361pt}{0.400pt}}
    \multiput(770.00,832.61)(12.965,0.447){3}{\rule{7.967pt}{0.108pt}}
    \multiput(770.00,831.17)(42.465,3.000){2}{\rule{3.983pt}{0.400pt}}
    \put(829,834.67){\rule{11.322pt}{0.400pt}}
    \multiput(829.00,834.17)(23.500,1.000){2}{\rule{5.661pt}{0.400pt}}
    \put(182,131){\makebox(0,0){$+$}}
    \put(241,131){\makebox(0,0){$+$}}
    \put(300,131){\makebox(0,0){$+$}}
    \put(417,131){\makebox(0,0){$+$}}
    \put(476,131){\makebox(0,0){$+$}}
    \put(488,131){\makebox(0,0){$+$}}
    \put(500,585){\makebox(0,0){$+$}}
    \put(511,657){\makebox(0,0){$+$}}
    \put(523,695){\makebox(0,0){$+$}}
    \put(535,728){\makebox(0,0){$+$}}
    \put(559,752){\makebox(0,0){$+$}}
    \put(582,780){\makebox(0,0){$+$}}
    \put(606,789){\makebox(0,0){$+$}}
    \put(653,812){\makebox(0,0){$+$}}
    \put(711,823){\makebox(0,0){$+$}}
    \put(770,832){\makebox(0,0){$+$}}
    \put(829,835){\makebox(0,0){$+$}}
    \put(876,836){\makebox(0,0){$+$}}
    \put(406,604){\makebox(0,0){$+$}}
    \put(182.0,131.0){\rule[-0.200pt]{73.715pt}{0.400pt}}
    \put(336,563){\makebox(0,0)[r]{\small{analytical}}}
    \put(356.0,563.0){\rule[-0.200pt]{24.090pt}{0.400pt}}
    \put(182,131){\usebox{\plotpoint}}
    \multiput(510.59,131.00)(0.485,39.014){11}{\rule{0.117pt}{29.357pt}}
    \multiput(509.17,131.00)(7.000,451.068){2}{\rule{0.400pt}{14.679pt}}
    \multiput(517.59,643.00)(0.485,4.459){11}{\rule{0.117pt}{3.471pt}}
    \multiput(516.17,643.00)(7.000,51.795){2}{\rule{0.400pt}{1.736pt}}
    \multiput(524.59,702.00)(0.485,2.018){11}{\rule{0.117pt}{1.643pt}}
    \multiput(523.17,702.00)(7.000,23.590){2}{\rule{0.400pt}{0.821pt}}
    \multiput(531.59,729.00)(0.488,1.088){13}{\rule{0.117pt}{0.950pt}}
    \multiput(530.17,729.00)(8.000,15.028){2}{\rule{0.400pt}{0.475pt}}
    \multiput(539.59,746.00)(0.485,0.874){11}{\rule{0.117pt}{0.786pt}}
    \multiput(538.17,746.00)(7.000,10.369){2}{\rule{0.400pt}{0.393pt}}
    \multiput(546.59,758.00)(0.485,0.645){11}{\rule{0.117pt}{0.614pt}}
    \multiput(545.17,758.00)(7.000,7.725){2}{\rule{0.400pt}{0.307pt}}
    \multiput(553.00,767.59)(0.492,0.485){11}{\rule{0.500pt}{0.117pt}}
    \multiput(553.00,766.17)(5.962,7.000){2}{\rule{0.250pt}{0.400pt}}
    \multiput(560.00,774.59)(0.581,0.482){9}{\rule{0.567pt}{0.116pt}}
    \multiput(560.00,773.17)(5.824,6.000){2}{\rule{0.283pt}{0.400pt}}
    \multiput(567.00,780.59)(0.710,0.477){7}{\rule{0.660pt}{0.115pt}}
    \multiput(567.00,779.17)(5.630,5.000){2}{\rule{0.330pt}{0.400pt}}
    \multiput(574.00,785.59)(0.710,0.477){7}{\rule{0.660pt}{0.115pt}}
    \multiput(574.00,784.17)(5.630,5.000){2}{\rule{0.330pt}{0.400pt}}
    \multiput(581.00,790.60)(0.920,0.468){5}{\rule{0.800pt}{0.113pt}}
    \multiput(581.00,789.17)(5.340,4.000){2}{\rule{0.400pt}{0.400pt}}
    \multiput(588.00,794.61)(1.579,0.447){3}{\rule{1.167pt}{0.108pt}}
    \multiput(588.00,793.17)(5.579,3.000){2}{\rule{0.583pt}{0.400pt}}
    \multiput(596.00,797.61)(1.355,0.447){3}{\rule{1.033pt}{0.108pt}}
    \multiput(596.00,796.17)(4.855,3.000){2}{\rule{0.517pt}{0.400pt}}
    \multiput(603.00,800.61)(1.355,0.447){3}{\rule{1.033pt}{0.108pt}}
    \multiput(603.00,799.17)(4.855,3.000){2}{\rule{0.517pt}{0.400pt}}
    \put(610,803.17){\rule{1.500pt}{0.400pt}}
    \multiput(610.00,802.17)(3.887,2.000){2}{\rule{0.750pt}{0.400pt}}
    \multiput(617.00,805.61)(1.355,0.447){3}{\rule{1.033pt}{0.108pt}}
    \multiput(617.00,804.17)(4.855,3.000){2}{\rule{0.517pt}{0.400pt}}
    \put(624,808.17){\rule{1.500pt}{0.400pt}}
    \multiput(624.00,807.17)(3.887,2.000){2}{\rule{0.750pt}{0.400pt}}
    \put(631,809.67){\rule{1.686pt}{0.400pt}}
    \multiput(631.00,809.17)(3.500,1.000){2}{\rule{0.843pt}{0.400pt}}
    \put(638,811.17){\rule{1.700pt}{0.400pt}}
    \multiput(638.00,810.17)(4.472,2.000){2}{\rule{0.850pt}{0.400pt}}
    \put(646,813.17){\rule{1.500pt}{0.400pt}}
    \multiput(646.00,812.17)(3.887,2.000){2}{\rule{0.750pt}{0.400pt}}
    \put(653,814.67){\rule{1.686pt}{0.400pt}}
    \multiput(653.00,814.17)(3.500,1.000){2}{\rule{0.843pt}{0.400pt}}
    \put(660,816.17){\rule{1.500pt}{0.400pt}}
    \multiput(660.00,815.17)(3.887,2.000){2}{\rule{0.750pt}{0.400pt}}
    \put(667,817.67){\rule{1.686pt}{0.400pt}}
    \multiput(667.00,817.17)(3.500,1.000){2}{\rule{0.843pt}{0.400pt}}
    \put(674,818.67){\rule{1.686pt}{0.400pt}}
    \multiput(674.00,818.17)(3.500,1.000){2}{\rule{0.843pt}{0.400pt}}
    \put(681,820.17){\rule{1.500pt}{0.400pt}}
    \multiput(681.00,819.17)(3.887,2.000){2}{\rule{0.750pt}{0.400pt}}
    \put(688,821.67){\rule{1.686pt}{0.400pt}}
    \multiput(688.00,821.17)(3.500,1.000){2}{\rule{0.843pt}{0.400pt}}
    \put(695,822.67){\rule{1.927pt}{0.400pt}}
    \multiput(695.00,822.17)(4.000,1.000){2}{\rule{0.964pt}{0.400pt}}
    \put(703,823.67){\rule{1.686pt}{0.400pt}}
    \multiput(703.00,823.17)(3.500,1.000){2}{\rule{0.843pt}{0.400pt}}
    \put(710,824.67){\rule{1.686pt}{0.400pt}}
    \multiput(710.00,824.17)(3.500,1.000){2}{\rule{0.843pt}{0.400pt}}
    \put(182.0,131.0){\rule[-0.200pt]{79.015pt}{0.400pt}}
    \put(724,825.67){\rule{1.686pt}{0.400pt}}
    \multiput(724.00,825.17)(3.500,1.000){2}{\rule{0.843pt}{0.400pt}}
    \put(731,826.67){\rule{1.686pt}{0.400pt}}
    \multiput(731.00,826.17)(3.500,1.000){2}{\rule{0.843pt}{0.400pt}}
    \put(738,827.67){\rule{1.686pt}{0.400pt}}
    \multiput(738.00,827.17)(3.500,1.000){2}{\rule{0.843pt}{0.400pt}}
    \put(717.0,826.0){\rule[-0.200pt]{1.686pt}{0.400pt}}
    \put(753,828.67){\rule{1.686pt}{0.400pt}}
    \multiput(753.00,828.17)(3.500,1.000){2}{\rule{0.843pt}{0.400pt}}
    \put(760,829.67){\rule{1.686pt}{0.400pt}}
    \multiput(760.00,829.17)(3.500,1.000){2}{\rule{0.843pt}{0.400pt}}
    \put(745.0,829.0){\rule[-0.200pt]{1.927pt}{0.400pt}}
    \put(774,830.67){\rule{1.686pt}{0.400pt}}
    \multiput(774.00,830.17)(3.500,1.000){2}{\rule{0.843pt}{0.400pt}}
    \put(767.0,831.0){\rule[-0.200pt]{1.686pt}{0.400pt}}
    \put(788,831.67){\rule{1.686pt}{0.400pt}}
    \multiput(788.00,831.17)(3.500,1.000){2}{\rule{0.843pt}{0.400pt}}
    \put(781.0,832.0){\rule[-0.200pt]{1.686pt}{0.400pt}}
    \put(810,832.67){\rule{1.686pt}{0.400pt}}
    \multiput(810.00,832.17)(3.500,1.000){2}{\rule{0.843pt}{0.400pt}}
    \put(795.0,833.0){\rule[-0.200pt]{3.613pt}{0.400pt}}
    \put(824,833.67){\rule{1.686pt}{0.400pt}}
    \multiput(824.00,833.17)(3.500,1.000){2}{\rule{0.843pt}{0.400pt}}
    \put(817.0,834.0){\rule[-0.200pt]{1.686pt}{0.400pt}}
    \put(852,834.67){\rule{1.686pt}{0.400pt}}
    \multiput(852.00,834.17)(3.500,1.000){2}{\rule{0.843pt}{0.400pt}}
    \put(831.0,835.0){\rule[-0.200pt]{5.059pt}{0.400pt}}
    \put(182,131){\makebox(0,0){$\bullet$}}
    \put(189,131){\makebox(0,0){$\bullet$}}
    \put(196,131){\makebox(0,0){$\bullet$}}
    \put(203,131){\makebox(0,0){$\bullet$}}
    \put(211,131){\makebox(0,0){$\bullet$}}
    \put(218,131){\makebox(0,0){$\bullet$}}
    \put(225,131){\makebox(0,0){$\bullet$}}
    \put(232,131){\makebox(0,0){$\bullet$}}
    \put(239,131){\makebox(0,0){$\bullet$}}
    \put(246,131){\makebox(0,0){$\bullet$}}
    \put(253,131){\makebox(0,0){$\bullet$}}
    \put(260,131){\makebox(0,0){$\bullet$}}
    \put(268,131){\makebox(0,0){$\bullet$}}
    \put(275,131){\makebox(0,0){$\bullet$}}
    \put(282,131){\makebox(0,0){$\bullet$}}
    \put(289,131){\makebox(0,0){$\bullet$}}
    \put(296,131){\makebox(0,0){$\bullet$}}
    \put(303,131){\makebox(0,0){$\bullet$}}
    \put(310,131){\makebox(0,0){$\bullet$}}
    \put(317,131){\makebox(0,0){$\bullet$}}
    \put(325,131){\makebox(0,0){$\bullet$}}
    \put(332,131){\makebox(0,0){$\bullet$}}
    \put(339,131){\makebox(0,0){$\bullet$}}
    \put(346,131){\makebox(0,0){$\bullet$}}
    \put(353,131){\makebox(0,0){$\bullet$}}
    \put(360,131){\makebox(0,0){$\bullet$}}
    \put(367,131){\makebox(0,0){$\bullet$}}
    \put(375,131){\makebox(0,0){$\bullet$}}
    \put(382,131){\makebox(0,0){$\bullet$}}
    \put(389,131){\makebox(0,0){$\bullet$}}
    \put(396,131){\makebox(0,0){$\bullet$}}
    \put(403,131){\makebox(0,0){$\bullet$}}
    \put(410,131){\makebox(0,0){$\bullet$}}
    \put(417,131){\makebox(0,0){$\bullet$}}
    \put(424,131){\makebox(0,0){$\bullet$}}
    \put(432,131){\makebox(0,0){$\bullet$}}
    \put(439,131){\makebox(0,0){$\bullet$}}
    \put(446,131){\makebox(0,0){$\bullet$}}
    \put(453,131){\makebox(0,0){$\bullet$}}
    \put(460,131){\makebox(0,0){$\bullet$}}
    \put(467,131){\makebox(0,0){$\bullet$}}
    \put(474,131){\makebox(0,0){$\bullet$}}
    \put(482,131){\makebox(0,0){$\bullet$}}
    \put(489,131){\makebox(0,0){$\bullet$}}
    \put(496,131){\makebox(0,0){$\bullet$}}
    \put(503,131){\makebox(0,0){$\bullet$}}
    \put(510,131){\makebox(0,0){$\bullet$}}
    \put(517,643){\makebox(0,0){$\bullet$}}
    \put(524,702){\makebox(0,0){$\bullet$}}
    \put(531,729){\makebox(0,0){$\bullet$}}
    \put(539,746){\makebox(0,0){$\bullet$}}
    \put(546,758){\makebox(0,0){$\bullet$}}
    \put(553,767){\makebox(0,0){$\bullet$}}
    \put(560,774){\makebox(0,0){$\bullet$}}
    \put(567,780){\makebox(0,0){$\bullet$}}
    \put(574,785){\makebox(0,0){$\bullet$}}
    \put(581,790){\makebox(0,0){$\bullet$}}
    \put(588,794){\makebox(0,0){$\bullet$}}
    \put(596,797){\makebox(0,0){$\bullet$}}
    \put(603,800){\makebox(0,0){$\bullet$}}
    \put(610,803){\makebox(0,0){$\bullet$}}
    \put(617,805){\makebox(0,0){$\bullet$}}
    \put(624,808){\makebox(0,0){$\bullet$}}
    \put(631,810){\makebox(0,0){$\bullet$}}
    \put(638,811){\makebox(0,0){$\bullet$}}
    \put(646,813){\makebox(0,0){$\bullet$}}
    \put(653,815){\makebox(0,0){$\bullet$}}
    \put(660,816){\makebox(0,0){$\bullet$}}
    \put(667,818){\makebox(0,0){$\bullet$}}
    \put(674,819){\makebox(0,0){$\bullet$}}
    \put(681,820){\makebox(0,0){$\bullet$}}
    \put(688,822){\makebox(0,0){$\bullet$}}
    \put(695,823){\makebox(0,0){$\bullet$}}
    \put(703,824){\makebox(0,0){$\bullet$}}
    \put(710,825){\makebox(0,0){$\bullet$}}
    \put(717,826){\makebox(0,0){$\bullet$}}
    \put(724,826){\makebox(0,0){$\bullet$}}
    \put(731,827){\makebox(0,0){$\bullet$}}
    \put(738,828){\makebox(0,0){$\bullet$}}
    \put(745,829){\makebox(0,0){$\bullet$}}
    \put(753,829){\makebox(0,0){$\bullet$}}
    \put(760,830){\makebox(0,0){$\bullet$}}
    \put(767,831){\makebox(0,0){$\bullet$}}
    \put(774,831){\makebox(0,0){$\bullet$}}
    \put(781,832){\makebox(0,0){$\bullet$}}
    \put(788,832){\makebox(0,0){$\bullet$}}
    \put(795,833){\makebox(0,0){$\bullet$}}
    \put(802,833){\makebox(0,0){$\bullet$}}
    \put(810,833){\makebox(0,0){$\bullet$}}
    \put(817,834){\makebox(0,0){$\bullet$}}
    \put(824,834){\makebox(0,0){$\bullet$}}
    \put(831,835){\makebox(0,0){$\bullet$}}
    \put(838,835){\makebox(0,0){$\bullet$}}
    \put(845,835){\makebox(0,0){$\bullet$}}
    \put(852,835){\makebox(0,0){$\bullet$}}
    \put(859,836){\makebox(0,0){$\bullet$}}
    \put(867,836){\makebox(0,0){$\bullet$}}
    \put(874,836){\makebox(0,0){$\bullet$}}
    \put(881,836){\makebox(0,0){$\bullet$}}
    \put(888,836){\makebox(0,0){$\bullet$}}
    \put(406,563){\makebox(0,0){$\bullet$}}
    \put(859.0,836.0){\rule[-0.200pt]{6.986pt}{0.400pt}}
    \put(182.0,131.0){\rule[-0.200pt]{0.400pt}{169.834pt}}
    \put(182.0,131.0){\rule[-0.200pt]{170.075pt}{0.400pt}}
    \put(888.0,131.0){\rule[-0.200pt]{0.400pt}{169.834pt}}
    \put(182.0,836.0){\rule[-0.200pt]{170.075pt}{0.400pt}}
  \end{picture}
  \caption{$|O_a|$ plotted against $\lambda$. The data points are
    obtained from a numerical simulation, the solid line is given by
    eq. (\ref{prediction}).}
  \label{critical}
\end{figure}

Figure~\ref{critical} shows the results of a simlulation
calculating $O$ as a function of $\lambda$ at steady state, and the
prediction of eq. (\ref{prediction}), which is in excellent agreement
with the data for $k=0.7$.
Up until $\lambda_c$, the system is in a symmetric (disordered) state
($O_a=0$) in which the opinions average out. At $\lambda_c$ the
system undergoes a critical symmetry breaking, and becomes quickly
completely polarized, either on a positive opinion
($O_a\approx{1}$) or a negative one ($O_infty\approx{1}$, the two
occurr with equal probability).

\section{Lattice model with local interaction}
The model presented in the previous section assumed that each
individual could interact indifferently with any other individual. In
this case, spontaneous symmetry breaking occurs for $\lambda_c=e/4$. A
richer model can be obtained by considering \emph{placed} individuals,
which interact with their neighbors. We ask whether in this case a
similar symmetry breaking transition occurs and what are its
characteristics. We consider a grid of $N\times{N}$ individuals where,
as in the previous model, the individual $(i,j)$ holds an opinion
$O_{ij}(t)$ at time $t$. The interaction of individual $(i,j)$ with
individual $(h,k)$ is given, as before, by
\begin{equation}
O_{i,j}(t+1) = \boxx{\lambda (O_{i,j} + \epsilon_t O_{h,k}(t))} 
\end{equation}
\begin{figure}[hbt]
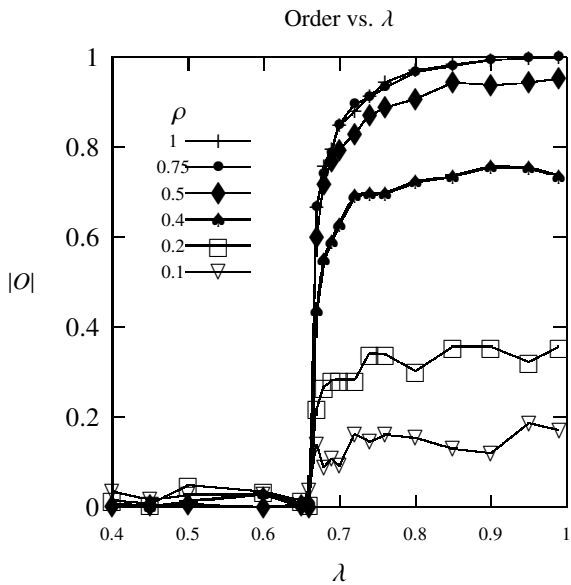

  \setlength{\unitlength}{0.240900pt}
  \ifx\plotpoint\undefined\newsavebox{\plotpoint}\fi
  \sbox{\plotpoint}{\rule[-0.200pt]{0.400pt}{0.400pt}}%

  \caption{$|O_a|$ plotted against $\lambda$ for a system with
    limited range interaction for different values of the localization
    paramter $\rho$.}
  \label{lim_O}
\end{figure}
(Note that we use the asymmetric version of the evolution equation: in
each interaction, only one individual change opinion; this choice
doesn't change the steady state of the interaction, although it
multiply by two the transition time.) However, unlike the previous
model, the individuals $(i,j)$ and $(h,k)$ can interact only if
$|i-h|\le{r}$ and $|j-k|\le{r}$, where $r$ is the \emph{range} of the
interaction. Note that if $r\ge{N}$ the system is again a fully
interconnected one. The parameter that characterizes te system is the
localization paramter $\rho=r/N$, which determine how far, relative to
the size of the lattice, can an individual interact.

The presence of a finite interaction range changes the characteristics
of the system, as shown in figure~\ref{lim_O},
in which $|O_a|$ is plotted against $\lambda$ for various values
of the interaction parameter. For $\rho>\rho_c\approx{1/2}$, the
behavior of the local interaction system is practically equivalent to
that of the fully connected, with a critical symmetry breaking
transition at $\lambda_c$.

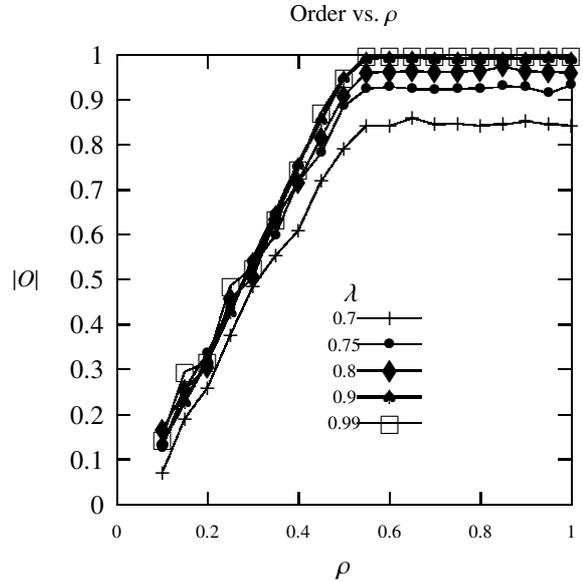
\begin{figure}[bth]
  \setlength{\unitlength}{0.240900pt}
  \ifx\plotpoint\undefined\newsavebox{\plotpoint}\fi
  \sbox{\plotpoint}{\rule[-0.200pt]{0.400pt}{0.400pt}}%
  \begin{picture}(960,960)(0,0)
    \sbox{\plotpoint}{\rule[-0.200pt]{0.400pt}{0.400pt}}%
    \put(182.0,131.0){\rule[-0.200pt]{4.818pt}{0.400pt}}
    \put(162,131){\makebox(0,0)[r]{ 0}}
    \put(868.0,131.0){\rule[-0.200pt]{4.818pt}{0.400pt}}
    \put(182.0,202.0){\rule[-0.200pt]{4.818pt}{0.400pt}}
    \put(162,202){\makebox(0,0)[r]{ 0.1}}
    \put(868.0,202.0){\rule[-0.200pt]{4.818pt}{0.400pt}}
    \put(182.0,272.0){\rule[-0.200pt]{4.818pt}{0.400pt}}
    \put(162,272){\makebox(0,0)[r]{ 0.2}}
    \put(868.0,272.0){\rule[-0.200pt]{4.818pt}{0.400pt}}
    \put(182.0,343.0){\rule[-0.200pt]{4.818pt}{0.400pt}}
    \put(162,343){\makebox(0,0)[r]{ 0.3}}
    \put(868.0,343.0){\rule[-0.200pt]{4.818pt}{0.400pt}}
    \put(182.0,413.0){\rule[-0.200pt]{4.818pt}{0.400pt}}
    \put(162,413){\makebox(0,0)[r]{ 0.4}}
    \put(868.0,413.0){\rule[-0.200pt]{4.818pt}{0.400pt}}
    \put(182.0,484.0){\rule[-0.200pt]{4.818pt}{0.400pt}}
    \put(162,484){\makebox(0,0)[r]{ 0.5}}
    \put(868.0,484.0){\rule[-0.200pt]{4.818pt}{0.400pt}}
    \put(182.0,554.0){\rule[-0.200pt]{4.818pt}{0.400pt}}
    \put(162,554){\makebox(0,0)[r]{ 0.6}}
    \put(868.0,554.0){\rule[-0.200pt]{4.818pt}{0.400pt}}
    \put(182.0,624.0){\rule[-0.200pt]{4.818pt}{0.400pt}}
    \put(162,624){\makebox(0,0)[r]{ 0.7}}
    \put(868.0,624.0){\rule[-0.200pt]{4.818pt}{0.400pt}}
    \put(182.0,695.0){\rule[-0.200pt]{4.818pt}{0.400pt}}
    \put(162,695){\makebox(0,0)[r]{ 0.8}}
    \put(868.0,695.0){\rule[-0.200pt]{4.818pt}{0.400pt}}
    \put(182.0,765.0){\rule[-0.200pt]{4.818pt}{0.400pt}}
    \put(162,765){\makebox(0,0)[r]{ 0.9}}
    \put(868.0,765.0){\rule[-0.200pt]{4.818pt}{0.400pt}}
    \put(182.0,836.0){\rule[-0.200pt]{4.818pt}{0.400pt}}
    \put(162,836){\makebox(0,0)[r]{ 1}}
    \put(868.0,836.0){\rule[-0.200pt]{4.818pt}{0.400pt}}
    \put(182.0,131.0){\rule[-0.200pt]{0.400pt}{4.818pt}}
    \put(182,90){\makebox(0,0){\scriptsize 0}}
    \put(182.0,816.0){\rule[-0.200pt]{0.400pt}{4.818pt}}
    \put(323.0,131.0){\rule[-0.200pt]{0.400pt}{4.818pt}}
    \put(323,90){\makebox(0,0){\scriptsize 0.2}}
    \put(323.0,816.0){\rule[-0.200pt]{0.400pt}{4.818pt}}
    \put(464.0,131.0){\rule[-0.200pt]{0.400pt}{4.818pt}}
    \put(464,90){\makebox(0,0){\scriptsize 0.4}}
    \put(464.0,816.0){\rule[-0.200pt]{0.400pt}{4.818pt}}
    \put(606.0,131.0){\rule[-0.200pt]{0.400pt}{4.818pt}}
    \put(606,90){\makebox(0,0){\scriptsize 0.6}}
    \put(606.0,816.0){\rule[-0.200pt]{0.400pt}{4.818pt}}
    \put(747.0,131.0){\rule[-0.200pt]{0.400pt}{4.818pt}}
    \put(747,90){\makebox(0,0){\scriptsize 0.8}}
    \put(747.0,816.0){\rule[-0.200pt]{0.400pt}{4.818pt}}
    \put(888.0,131.0){\rule[-0.200pt]{0.400pt}{4.818pt}}
    \put(888,90){\makebox(0,0){\scriptsize 1}}
    \put(888.0,816.0){\rule[-0.200pt]{0.400pt}{4.818pt}}
    \put(182.0,131.0){\rule[-0.200pt]{0.400pt}{169.834pt}}
    \put(182.0,131.0){\rule[-0.200pt]{170.075pt}{0.400pt}}
    \put(888.0,131.0){\rule[-0.200pt]{0.400pt}{169.834pt}}
    \put(182.0,836.0){\rule[-0.200pt]{170.075pt}{0.400pt}}
    \put(41,483){\makebox(0,0){\small{$|O|$}}}
    \put(535,29){\makebox(0,0){$\rho$}}
    \put(535,898){\makebox(0,0){\small{Order vs.\ $\rho$}}}
    \put(546,463){\makebox(0,0){$\lambda$}}
    \put(536,423){\makebox(0,0){\scriptsize 0.7}}
    \put(556.0,423.0){\rule[-0.200pt]{24.090pt}{0.400pt}}
    \put(253,181){\usebox{\plotpoint}}
    \multiput(253.58,181.00)(0.498,1.220){67}{\rule{0.120pt}{1.071pt}}
    \multiput(252.17,181.00)(35.000,82.776){2}{\rule{0.400pt}{0.536pt}}
    \multiput(288.58,266.00)(0.498,0.686){67}{\rule{0.120pt}{0.649pt}}
    \multiput(287.17,266.00)(35.000,46.654){2}{\rule{0.400pt}{0.324pt}}
    \multiput(323.58,314.00)(0.498,1.158){69}{\rule{0.120pt}{1.022pt}}
    \multiput(322.17,314.00)(36.000,80.878){2}{\rule{0.400pt}{0.511pt}}
    \multiput(359.58,397.00)(0.498,1.091){67}{\rule{0.120pt}{0.969pt}}
    \multiput(358.17,397.00)(35.000,73.990){2}{\rule{0.400pt}{0.484pt}}
    \multiput(394.58,473.00)(0.498,0.686){67}{\rule{0.120pt}{0.649pt}}
    \multiput(393.17,473.00)(35.000,46.654){2}{\rule{0.400pt}{0.324pt}}
    \multiput(429.58,521.00)(0.498,0.557){67}{\rule{0.120pt}{0.546pt}}
    \multiput(428.17,521.00)(35.000,37.867){2}{\rule{0.400pt}{0.273pt}}
    \multiput(464.58,560.00)(0.498,1.088){69}{\rule{0.120pt}{0.967pt}}
    \multiput(463.17,560.00)(36.000,75.994){2}{\rule{0.400pt}{0.483pt}}
    \multiput(500.58,638.00)(0.498,0.730){67}{\rule{0.120pt}{0.683pt}}
    \multiput(499.17,638.00)(35.000,49.583){2}{\rule{0.400pt}{0.341pt}}
    \multiput(535.58,689.00)(0.498,0.513){67}{\rule{0.120pt}{0.511pt}}
    \multiput(534.17,689.00)(35.000,34.939){2}{\rule{0.400pt}{0.256pt}}
    \multiput(606.00,725.58)(1.487,0.492){21}{\rule{1.267pt}{0.119pt}}
    \multiput(606.00,724.17)(32.371,12.000){2}{\rule{0.633pt}{0.400pt}}
    \multiput(641.00,735.92)(1.798,-0.491){17}{\rule{1.500pt}{0.118pt}}
    \multiput(641.00,736.17)(31.887,-10.000){2}{\rule{0.750pt}{0.400pt}}
    \put(676,726.67){\rule{8.672pt}{0.400pt}}
    \multiput(676.00,726.17)(18.000,1.000){2}{\rule{4.336pt}{0.400pt}}
    \multiput(712.00,726.95)(7.607,-0.447){3}{\rule{4.767pt}{0.108pt}}
    \multiput(712.00,727.17)(25.107,-3.000){2}{\rule{2.383pt}{0.400pt}}
    \put(747,725.17){\rule{7.100pt}{0.400pt}}
    \multiput(747.00,724.17)(20.264,2.000){2}{\rule{3.550pt}{0.400pt}}
    \multiput(782.00,727.59)(3.827,0.477){7}{\rule{2.900pt}{0.115pt}}
    \multiput(782.00,726.17)(28.981,5.000){2}{\rule{1.450pt}{0.400pt}}
    \multiput(817.00,730.93)(3.938,-0.477){7}{\rule{2.980pt}{0.115pt}}
    \multiput(817.00,731.17)(29.815,-5.000){2}{\rule{1.490pt}{0.400pt}}
    \put(853,725.17){\rule{7.100pt}{0.400pt}}
    \multiput(853.00,726.17)(20.264,-2.000){2}{\rule{3.550pt}{0.400pt}}
    \put(253,181){\makebox(0,0){$+$}}
    \put(288,266){\makebox(0,0){$+$}}
    \put(323,314){\makebox(0,0){$+$}}
    \put(359,397){\makebox(0,0){$+$}}
    \put(394,473){\makebox(0,0){$+$}}
    \put(429,521){\makebox(0,0){$+$}}
    \put(464,560){\makebox(0,0){$+$}}
    \put(500,638){\makebox(0,0){$+$}}
    \put(535,689){\makebox(0,0){$+$}}
    \put(570,725){\makebox(0,0){$+$}}
    \put(606,725){\makebox(0,0){$+$}}
    \put(641,737){\makebox(0,0){$+$}}
    \put(676,727){\makebox(0,0){$+$}}
    \put(712,728){\makebox(0,0){$+$}}
    \put(747,725){\makebox(0,0){$+$}}
    \put(782,727){\makebox(0,0){$+$}}
    \put(817,732){\makebox(0,0){$+$}}
    \put(853,727){\makebox(0,0){$+$}}
    \put(888,725){\makebox(0,0){$+$}}
    \put(606,423){\makebox(0,0){$+$}}
    \put(570.0,725.0){\rule[-0.200pt]{8.672pt}{0.400pt}}
    \put(536,382){\makebox(0,0){\scriptsize 0.75}}
    \put(556.0,382.0){\rule[-0.200pt]{24.090pt}{0.400pt}}
    \put(253,220){\usebox{\plotpoint}}
    \multiput(253.58,220.00)(0.498,1.322){67}{\rule{0.120pt}{1.151pt}}
    \multiput(252.17,220.00)(35.000,89.610){2}{\rule{0.400pt}{0.576pt}}
    \multiput(288.58,312.00)(0.498,0.816){67}{\rule{0.120pt}{0.751pt}}
    \multiput(287.17,312.00)(35.000,55.440){2}{\rule{0.400pt}{0.376pt}}
    \multiput(323.58,369.00)(0.498,0.976){69}{\rule{0.120pt}{0.878pt}}
    \multiput(322.17,369.00)(36.000,68.178){2}{\rule{0.400pt}{0.439pt}}
    \multiput(359.58,439.00)(0.498,0.946){67}{\rule{0.120pt}{0.854pt}}
    \multiput(358.17,439.00)(35.000,64.227){2}{\rule{0.400pt}{0.427pt}}
    \multiput(394.58,505.00)(0.498,0.686){67}{\rule{0.120pt}{0.649pt}}
    \multiput(393.17,505.00)(35.000,46.654){2}{\rule{0.400pt}{0.324pt}}
    \multiput(429.58,553.00)(0.498,1.206){67}{\rule{0.120pt}{1.060pt}}
    \multiput(428.17,553.00)(35.000,81.800){2}{\rule{0.400pt}{0.530pt}}
    \multiput(464.58,637.00)(0.498,0.625){69}{\rule{0.120pt}{0.600pt}}
    \multiput(463.17,637.00)(36.000,43.755){2}{\rule{0.400pt}{0.300pt}}
    \multiput(500.58,682.00)(0.498,1.062){67}{\rule{0.120pt}{0.946pt}}
    \multiput(499.17,682.00)(35.000,72.037){2}{\rule{0.400pt}{0.473pt}}
    \multiput(535.00,756.58)(0.649,0.497){51}{\rule{0.619pt}{0.120pt}}
    \multiput(535.00,755.17)(33.716,27.000){2}{\rule{0.309pt}{0.400pt}}
    \multiput(570.00,783.61)(7.830,0.447){3}{\rule{4.900pt}{0.108pt}}
    \multiput(570.00,782.17)(25.830,3.000){2}{\rule{2.450pt}{0.400pt}}
    \multiput(606.00,784.94)(5.014,-0.468){5}{\rule{3.600pt}{0.113pt}}
    \multiput(606.00,785.17)(27.528,-4.000){2}{\rule{1.800pt}{0.400pt}}
    \put(641,780.67){\rule{8.432pt}{0.400pt}}
    \multiput(641.00,781.17)(17.500,-1.000){2}{\rule{4.216pt}{0.400pt}}
    \put(676,780.67){\rule{8.672pt}{0.400pt}}
    \multiput(676.00,780.17)(18.000,1.000){2}{\rule{4.336pt}{0.400pt}}
    \put(712,781.67){\rule{8.432pt}{0.400pt}}
    \multiput(712.00,781.17)(17.500,1.000){2}{\rule{4.216pt}{0.400pt}}
    \multiput(747.00,783.60)(5.014,0.468){5}{\rule{3.600pt}{0.113pt}}
    \multiput(747.00,782.17)(27.528,4.000){2}{\rule{1.800pt}{0.400pt}}
    \put(782,785.17){\rule{7.100pt}{0.400pt}}
    \multiput(782.00,786.17)(20.264,-2.000){2}{\rule{3.550pt}{0.400pt}}
    \multiput(817.00,783.93)(2.067,-0.489){15}{\rule{1.700pt}{0.118pt}}
    \multiput(817.00,784.17)(32.472,-9.000){2}{\rule{0.850pt}{0.400pt}}
    \multiput(853.00,776.58)(1.369,0.493){23}{\rule{1.177pt}{0.119pt}}
    \multiput(853.00,775.17)(32.557,13.000){2}{\rule{0.588pt}{0.400pt}}
    \put(253,220){\makebox(0,0){$\bullet$}}
    \put(288,312){\makebox(0,0){$\bullet$}}
    \put(323,369){\makebox(0,0){$\bullet$}}
    \put(359,439){\makebox(0,0){$\bullet$}}
    \put(394,505){\makebox(0,0){$\bullet$}}
    \put(429,553){\makebox(0,0){$\bullet$}}
    \put(464,637){\makebox(0,0){$\bullet$}}
    \put(500,682){\makebox(0,0){$\bullet$}}
    \put(535,756){\makebox(0,0){$\bullet$}}
    \put(570,783){\makebox(0,0){$\bullet$}}
    \put(606,786){\makebox(0,0){$\bullet$}}
    \put(641,782){\makebox(0,0){$\bullet$}}
    \put(676,781){\makebox(0,0){$\bullet$}}
    \put(712,782){\makebox(0,0){$\bullet$}}
    \put(747,783){\makebox(0,0){$\bullet$}}
    \put(782,787){\makebox(0,0){$\bullet$}}
    \put(817,785){\makebox(0,0){$\bullet$}}
    \put(853,776){\makebox(0,0){$\bullet$}}
    \put(888,789){\makebox(0,0){$\bullet$}}
    \put(606,382){\makebox(0,0){$\bullet$}}
    \put(536,341){\makebox(0,0){\scriptsize 0.8}}
    \put(556.0,341.0){\rule[-0.200pt]{24.090pt}{0.400pt}}
    \put(253,248){\usebox{\plotpoint}}
    \multiput(253.58,248.00)(0.498,0.903){67}{\rule{0.120pt}{0.820pt}}
    \multiput(252.17,248.00)(35.000,61.298){2}{\rule{0.400pt}{0.410pt}}
    \multiput(288.58,311.00)(0.498,0.513){67}{\rule{0.120pt}{0.511pt}}
    \multiput(287.17,311.00)(35.000,34.939){2}{\rule{0.400pt}{0.256pt}}
    \multiput(323.58,347.00)(0.498,1.509){69}{\rule{0.120pt}{1.300pt}}
    \multiput(322.17,347.00)(36.000,105.302){2}{\rule{0.400pt}{0.650pt}}
    \multiput(359.00,455.58)(0.546,0.497){61}{\rule{0.538pt}{0.120pt}}
    \multiput(359.00,454.17)(33.884,32.000){2}{\rule{0.269pt}{0.400pt}}
    \multiput(394.58,487.00)(0.498,1.293){67}{\rule{0.120pt}{1.129pt}}
    \multiput(393.17,487.00)(35.000,87.658){2}{\rule{0.400pt}{0.564pt}}
    \multiput(429.58,577.00)(0.498,0.845){67}{\rule{0.120pt}{0.774pt}}
    \multiput(428.17,577.00)(35.000,57.393){2}{\rule{0.400pt}{0.387pt}}
    \multiput(464.58,636.00)(0.498,0.976){69}{\rule{0.120pt}{0.878pt}}
    \multiput(463.17,636.00)(36.000,68.178){2}{\rule{0.400pt}{0.439pt}}
    \multiput(500.58,706.00)(0.498,0.932){67}{\rule{0.120pt}{0.843pt}}
    \multiput(499.17,706.00)(35.000,63.251){2}{\rule{0.400pt}{0.421pt}}
    \multiput(535.58,771.00)(0.498,0.528){67}{\rule{0.120pt}{0.523pt}}
    \multiput(534.17,771.00)(35.000,35.915){2}{\rule{0.400pt}{0.261pt}}
    \put(570,807.67){\rule{8.672pt}{0.400pt}}
    \multiput(570.00,807.17)(18.000,1.000){2}{\rule{4.336pt}{0.400pt}}
    \put(606,808.67){\rule{8.432pt}{0.400pt}}
    \multiput(606.00,808.17)(17.500,1.000){2}{\rule{4.216pt}{0.400pt}}
    \put(641,808.67){\rule{8.432pt}{0.400pt}}
    \multiput(641.00,809.17)(17.500,-1.000){2}{\rule{4.216pt}{0.400pt}}
    \put(712,809.17){\rule{7.100pt}{0.400pt}}
    \multiput(712.00,808.17)(20.264,2.000){2}{\rule{3.550pt}{0.400pt}}
    \multiput(747.00,811.59)(2.628,0.485){11}{\rule{2.100pt}{0.117pt}}
    \multiput(747.00,810.17)(30.641,7.000){2}{\rule{1.050pt}{0.400pt}}
    \multiput(782.00,816.93)(2.277,-0.488){13}{\rule{1.850pt}{0.117pt}}
    \multiput(782.00,817.17)(31.160,-8.000){2}{\rule{0.925pt}{0.400pt}}
    \put(817,808.67){\rule{8.672pt}{0.400pt}}
    \multiput(817.00,809.17)(18.000,-1.000){2}{\rule{4.336pt}{0.400pt}}
    \put(853,807.67){\rule{8.432pt}{0.400pt}}
    \multiput(853.00,808.17)(17.500,-1.000){2}{\rule{4.216pt}{0.400pt}}
    \put(253,248){\makebox(0,0){$\blacklozenge$}}
    \put(288,311){\makebox(0,0){$\blacklozenge$}}
    \put(323,347){\makebox(0,0){$\blacklozenge$}}
    \put(359,455){\makebox(0,0){$\blacklozenge$}}
    \put(394,487){\makebox(0,0){$\blacklozenge$}}
    \put(429,577){\makebox(0,0){$\blacklozenge$}}
    \put(464,636){\makebox(0,0){$\blacklozenge$}}
    \put(500,706){\makebox(0,0){$\blacklozenge$}}
    \put(535,771){\makebox(0,0){$\blacklozenge$}}
    \put(570,808){\makebox(0,0){$\blacklozenge$}}
    \put(606,809){\makebox(0,0){$\blacklozenge$}}
    \put(641,810){\makebox(0,0){$\blacklozenge$}}
    \put(676,809){\makebox(0,0){$\blacklozenge$}}
    \put(712,809){\makebox(0,0){$\blacklozenge$}}
    \put(747,811){\makebox(0,0){$\blacklozenge$}}
    \put(782,818){\makebox(0,0){$\blacklozenge$}}
    \put(817,810){\makebox(0,0){$\blacklozenge$}}
    \put(853,809){\makebox(0,0){$\blacklozenge$}}
    \put(888,808){\makebox(0,0){$\blacklozenge$}}
    \put(606,341){\makebox(0,0){$\blacklozenge$}}
    \put(676.0,809.0){\rule[-0.200pt]{8.672pt}{0.400pt}}
    \sbox{\plotpoint}{\rule[-0.400pt]{0.800pt}{0.800pt}}%
    \sbox{\plotpoint}{\rule[-0.200pt]{0.400pt}{0.400pt}}%
    \put(536,300){\makebox(0,0){\scriptsize 0.9}}
    \sbox{\plotpoint}{\rule[-0.400pt]{0.800pt}{0.800pt}}%
    \put(556.0,300.0){\rule[-0.400pt]{24.090pt}{0.800pt}}
    \put(253,229){\usebox{\plotpoint}}
    \multiput(254.41,229.00)(0.503,0.921){63}{\rule{0.121pt}{1.663pt}}
    \multiput(251.34,229.00)(35.000,60.549){2}{\rule{0.800pt}{0.831pt}}
    \multiput(289.41,293.00)(0.503,0.921){63}{\rule{0.121pt}{1.663pt}}
    \multiput(286.34,293.00)(35.000,60.549){2}{\rule{0.800pt}{0.831pt}}
    \multiput(324.41,357.00)(0.503,1.079){65}{\rule{0.121pt}{1.911pt}}
    \multiput(321.34,357.00)(36.000,73.033){2}{\rule{0.800pt}{0.956pt}}
    \multiput(360.41,434.00)(0.503,1.184){63}{\rule{0.121pt}{2.074pt}}
    \multiput(357.34,434.00)(35.000,77.695){2}{\rule{0.800pt}{1.037pt}}
    \multiput(395.41,516.00)(0.503,1.052){63}{\rule{0.121pt}{1.869pt}}
    \multiput(392.34,516.00)(35.000,69.122){2}{\rule{0.800pt}{0.934pt}}
    \multiput(430.41,589.00)(0.503,1.096){63}{\rule{0.121pt}{1.937pt}}
    \multiput(427.34,589.00)(35.000,71.979){2}{\rule{0.800pt}{0.969pt}}
    \multiput(465.41,665.00)(0.503,0.980){65}{\rule{0.121pt}{1.756pt}}
    \multiput(462.34,665.00)(36.000,66.356){2}{\rule{0.800pt}{0.878pt}}
    \multiput(501.41,735.00)(0.503,0.936){63}{\rule{0.121pt}{1.686pt}}
    \multiput(498.34,735.00)(35.000,61.501){2}{\rule{0.800pt}{0.843pt}}
    \multiput(535.00,801.41)(0.563,0.503){55}{\rule{1.103pt}{0.121pt}}
    \multiput(535.00,798.34)(32.710,31.000){2}{\rule{0.552pt}{0.800pt}}
    \put(570,829.84){\rule{8.672pt}{0.800pt}}
    \multiput(570.00,829.34)(18.000,1.000){2}{\rule{4.336pt}{0.800pt}}
    \put(641,829.84){\rule{8.432pt}{0.800pt}}
    \multiput(641.00,830.34)(17.500,-1.000){2}{\rule{4.216pt}{0.800pt}}
    \put(676,828.84){\rule{8.672pt}{0.800pt}}
    \multiput(676.00,829.34)(18.000,-1.000){2}{\rule{4.336pt}{0.800pt}}
    \put(712,828.84){\rule{8.432pt}{0.800pt}}
    \multiput(712.00,828.34)(17.500,1.000){2}{\rule{4.216pt}{0.800pt}}
    \put(606.0,832.0){\rule[-0.400pt]{8.431pt}{0.800pt}}
    \put(782,828.84){\rule{8.432pt}{0.800pt}}
    \multiput(782.00,829.34)(17.500,-1.000){2}{\rule{4.216pt}{0.800pt}}
    \put(817,829.34){\rule{8.672pt}{0.800pt}}
    \multiput(817.00,828.34)(18.000,2.000){2}{\rule{4.336pt}{0.800pt}}
    \put(853,829.84){\rule{8.432pt}{0.800pt}}
    \multiput(853.00,830.34)(17.500,-1.000){2}{\rule{4.216pt}{0.800pt}}
    \put(253,229){\makebox(0,0){$\spadesuit$}}
    \put(288,293){\makebox(0,0){$\spadesuit$}}
    \put(323,357){\makebox(0,0){$\spadesuit$}}
    \put(359,434){\makebox(0,0){$\spadesuit$}}
    \put(394,516){\makebox(0,0){$\spadesuit$}}
    \put(429,589){\makebox(0,0){$\spadesuit$}}
    \put(464,665){\makebox(0,0){$\spadesuit$}}
    \put(500,735){\makebox(0,0){$\spadesuit$}}
    \put(535,800){\makebox(0,0){$\spadesuit$}}
    \put(570,831){\makebox(0,0){$\spadesuit$}}
    \put(606,832){\makebox(0,0){$\spadesuit$}}
    \put(641,832){\makebox(0,0){$\spadesuit$}}
    \put(676,831){\makebox(0,0){$\spadesuit$}}
    \put(712,830){\makebox(0,0){$\spadesuit$}}
    \put(747,831){\makebox(0,0){$\spadesuit$}}
    \put(782,831){\makebox(0,0){$\spadesuit$}}
    \put(817,830){\makebox(0,0){$\spadesuit$}}
    \put(853,832){\makebox(0,0){$\spadesuit$}}
    \put(888,831){\makebox(0,0){$\spadesuit$}}
    \put(606,300){\makebox(0,0){$\spadesuit$}}
    \put(747.0,831.0){\rule[-0.400pt]{8.431pt}{0.800pt}}
    \sbox{\plotpoint}{\rule[-0.200pt]{0.400pt}{0.400pt}}%
    \put(536,259){\makebox(0,0){\scriptsize 0.99}}
    \put(556.0,259.0){\rule[-0.200pt]{24.090pt}{0.400pt}}
    \put(253,235){\usebox{\plotpoint}}
    \multiput(253.58,235.00)(0.498,1.509){67}{\rule{0.120pt}{1.300pt}}
    \multiput(252.17,235.00)(35.000,102.302){2}{\rule{0.400pt}{0.650pt}}
    \multiput(288.00,340.58)(1.105,0.494){29}{\rule{0.975pt}{0.119pt}}
    \multiput(288.00,339.17)(32.976,16.000){2}{\rule{0.488pt}{0.400pt}}
    \multiput(323.58,356.00)(0.498,1.663){69}{\rule{0.120pt}{1.422pt}}
    \multiput(322.17,356.00)(36.000,116.048){2}{\rule{0.400pt}{0.711pt}}
    \multiput(359.00,475.58)(0.625,0.497){53}{\rule{0.600pt}{0.120pt}}
    \multiput(359.00,474.17)(33.755,28.000){2}{\rule{0.300pt}{0.400pt}}
    \multiput(394.58,503.00)(0.498,1.091){67}{\rule{0.120pt}{0.969pt}}
    \multiput(393.17,503.00)(35.000,73.990){2}{\rule{0.400pt}{0.484pt}}
    \multiput(429.58,579.00)(0.498,1.134){67}{\rule{0.120pt}{1.003pt}}
    \multiput(428.17,579.00)(35.000,76.919){2}{\rule{0.400pt}{0.501pt}}
    \multiput(464.58,658.00)(0.498,1.228){69}{\rule{0.120pt}{1.078pt}}
    \multiput(463.17,658.00)(36.000,85.763){2}{\rule{0.400pt}{0.539pt}}
    \multiput(500.58,746.00)(0.498,0.788){67}{\rule{0.120pt}{0.729pt}}
    \multiput(499.17,746.00)(35.000,53.488){2}{\rule{0.400pt}{0.364pt}}
    \multiput(535.00,801.58)(0.499,0.498){67}{\rule{0.500pt}{0.120pt}}
    \multiput(535.00,800.17)(33.962,35.000){2}{\rule{0.250pt}{0.400pt}}
    \put(253,235){\raisebox{-.8pt}{\makebox(0,0){$\Box$}}}
    \put(288,340){\raisebox{-.8pt}{\makebox(0,0){$\Box$}}}
    \put(323,356){\raisebox{-.8pt}{\makebox(0,0){$\Box$}}}
    \put(359,475){\raisebox{-.8pt}{\makebox(0,0){$\Box$}}}
    \put(394,503){\raisebox{-.8pt}{\makebox(0,0){$\Box$}}}
    \put(429,579){\raisebox{-.8pt}{\makebox(0,0){$\Box$}}}
    \put(464,658){\raisebox{-.8pt}{\makebox(0,0){$\Box$}}}
    \put(500,746){\raisebox{-.8pt}{\makebox(0,0){$\Box$}}}
    \put(535,801){\raisebox{-.8pt}{\makebox(0,0){$\Box$}}}
    \put(570,836){\raisebox{-.8pt}{\makebox(0,0){$\Box$}}}
    \put(606,836){\raisebox{-.8pt}{\makebox(0,0){$\Box$}}}
    \put(641,836){\raisebox{-.8pt}{\makebox(0,0){$\Box$}}}
    \put(676,836){\raisebox{-.8pt}{\makebox(0,0){$\Box$}}}
    \put(712,836){\raisebox{-.8pt}{\makebox(0,0){$\Box$}}}
    \put(747,836){\raisebox{-.8pt}{\makebox(0,0){$\Box$}}}
    \put(782,836){\raisebox{-.8pt}{\makebox(0,0){$\Box$}}}
    \put(817,836){\raisebox{-.8pt}{\makebox(0,0){$\Box$}}}
    \put(853,836){\raisebox{-.8pt}{\makebox(0,0){$\Box$}}}
    \put(888,836){\raisebox{-.8pt}{\makebox(0,0){$\Box$}}}
    \put(606,259){\raisebox{-.8pt}{\makebox(0,0){$\Box$}}}
    \put(570.0,836.0){\rule[-0.200pt]{76.606pt}{0.400pt}}
    \put(182.0,131.0){\rule[-0.200pt]{0.400pt}{169.834pt}}
    \put(182.0,131.0){\rule[-0.200pt]{170.075pt}{0.400pt}}
    \put(888.0,131.0){\rule[-0.200pt]{0.400pt}{169.834pt}}
    \put(182.0,836.0){\rule[-0.200pt]{170.075pt}{0.400pt}}
  \end{picture}
  \caption{$|O_a|$ plotted against $\rho$ for a system with
    limited range interaction for different values of the conviction
    parameter $\lambda$ (in the range $[\lambda_c,1)$.}
    \label{lim_rho}
\end{figure}

\begin{figure}[tbh]
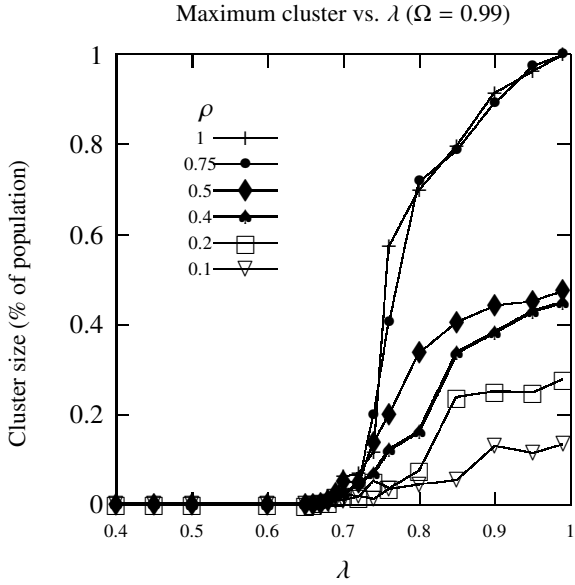

  \setlength{\unitlength}{0.240900pt}
  \ifx\plotpoint\undefined\newsavebox{\plotpoint}\fi
  \sbox{\plotpoint}{\rule[-0.200pt]{0.400pt}{0.400pt}}%

  \caption{Maximum cluster size as a function of the conviction
    parameter for various values of the localization parameter
    and $\Omega=0.99$.}
  \label{clst_O}
\end{figure}

For $\rho<\rho_c$, the symmetry still break for $\lambda=\lambda_c$,
but the order paramter doesn't reach the value $1$, stabilizing upon a
value that depends on $\rho$. For $\rho<\rho_c$, this value decreases
quite rapidly, as shown in figure~\ref{lim_rho}, which shows the value
of $|O_a|$ for various $\lambda>\lambda_c$ as a function of
$\rho$.
These lower values of the order parameter correspond to the
coexistence of various groups of individuals of different opinion.
This is shown in figure\ref{clst_O}, which shows the results of a
percolation study (see \cite{chandra:12} for a similar study on the
full interaction model).
Given an edge value $\Omega$, we consider a cluster as a group of
adyacent (4-neighborhood) individuals that have opinion
$O_{ij}>\Omega$ or $O_{ij}<\Omega$. The measure is significant mainly
for extreme opinions, that is, for $\Omega\approx{1}$.

Below the critical conviction $\lambda_c$, the system is in disorder,
and there is practically no formation of clusters. As $\lambda$
increases beyond $\lambda_c$, we see the formation of groups with
mutually reinforcing opinions, but the locality of the interaction
permits the creation of stable solutions with groups holding different
opinions.  The same conclusion can be reached by looking at the
formation of clusters as a function of the localization parameter
(figure~\ref{clst_rho}).
\begin{figure}[thb]
  \setlength{\unitlength}{0.240900pt}
  \ifx\plotpoint\undefined\newsavebox{\plotpoint}\fi
  \sbox{\plotpoint}{\rule[-0.200pt]{0.400pt}{0.400pt}}%

  \caption{Maximum cluster size as a function of the localization
    parameter $\rho$ various values of the conviction parameter and
    $\Omega=0.99$.}
  \label{clst_rho}
\end{figure}
From these results we observe what appears to be a more gradual
transition. For $\lambda\approx{1}$ we assist to the sudden break of
the symmetry around a critical point $\rho_c\approx{0.55}$. For
$\lambda\approx\lambda_c$ we are in the pre-percolation area
(cf. figure~\ref{clst_O}), and the clusters of opinion are consistently
small. In an intermediate area ($\lambda\approx{0.75}$ there appears
to be a phase of instability, in which the formation of clusters
varies widely (it must be noted that this is the area in which the
variance of the cluster size is higher).

We can have a little insight into the behavior of the system by
considering an extremely simplified case, that of a one-dimensional
continuous system in which the opinion of the individual $x$ at time
$t$ is a continuous function of $x$ $O(x;t)$, $x\in[-1,1]$ (we still
assume that time is discrete, so no continuity in time can not be
imposed). Consider a stable configuration, $O(x)$, for
$t\rightarrow\infty$, and assume that the individual at the origin has
opinion $O(0)=1$. We are asking whether the configuration with a
single cluster, that is, the configuration $O(x)=1$ is a stable
one. If $r$ is the interaction radius (note that the interaction span
is equal to $2r$ and the individual space has size two, so that
$\rho=r$), we can write the equilibrium as
\begin{equation}
\begin{array}{l}
\displaystyle O(0) = 1 \\
\displaystyle O(x) - \boxx{\lambda O(x) + \lambda\langle\epsilon\rangle\int_{x-r}^{x+r}\!\!\!\!\!\!\!\!\!O(u)\,du} = 0
\end{array}
\end{equation}
the solution $O(x)=1$ is stable if and only if 
\begin{equation}
\lambda + \lambda\langle\epsilon\rangle\int_{x-r}^{x+r}\!\!\!\!\!\!\!\!du =
\lambda (1 + 2 r \langle\epsilon\rangle) = \lambda (1+r) > 1
\end{equation}
that is, if $r>1/(1+\lambda)$. This would give us, for $\lambda=1$
point, that the formation of complete clusters is possible only for
$\rho>0.5$, vix. $\rho_c=0.5$, a bit less than the value that we
observe in the numerical simulations. We argue that this discrepancy
might be due to corner effects in the two-dimensional grid: the
two-dimensional equivalent of our one-dimensional analysis would be a
layout with radial symmetry, while the numerical analysis was carried
out on a square grid.

\section{Intreraction with recommendation}
So far we have dealt with the free exchange of opinion among
individual, without the introduction of a recommender. Here we shall
consider a very simple form of recommendation, but which should be
enough to capture the essential characteristics of this kind of
systems. Given an individual $O_i$ that is about to interact, we model
the recommendation by looking for the individual $O_{r(i)}$ closer to
$O_i$, that is
\begin{equation}
\label{recom1}
r(i) = \mbox{arg} \min_{j} |O(j)-O(i)|
\end{equation}
The value $O_{r(i)}$ is the recommendation made to the individual $i$,
which interacts with it using the standard interaction model of
eq. (\ref{model1}). When individual $i$ is about to interact, it will
choose to interact with the recommender system with probability $p$
and with an individual in its neighborhood with probability $1-p$. We
have found that the results are quite stable even for rather low
values of $p$, so we have set $p=0.05$, meaning that 5\% of the
interactions will be with the recommeder system.

\begin{figure}[bht]
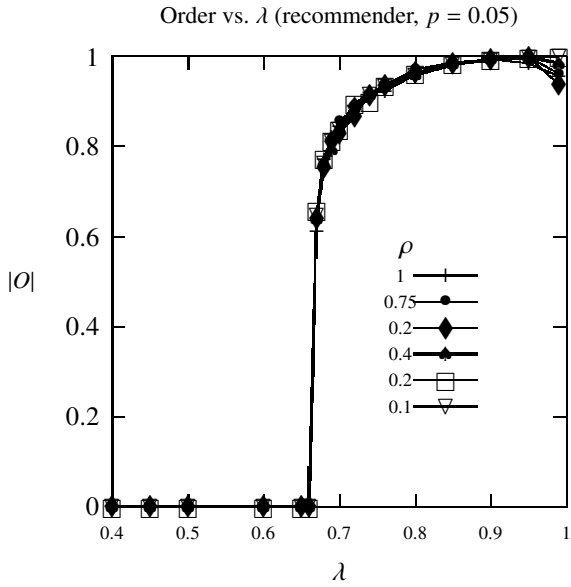

  \setlength{\unitlength}{0.240900pt}
  \ifx\plotpoint\undefined\newsavebox{\plotpoint}\fi
  \sbox{\plotpoint}{\rule[-0.200pt]{0.400pt}{0.400pt}}%

  \caption{$|O_a|$ plotted against $\lambda$ for a system with
    limited range interaction for different values of the localization
    paramter $\rho$ and a recommender system.}
  \label{lim_R}
\end{figure}

\begin{figure}[bht]
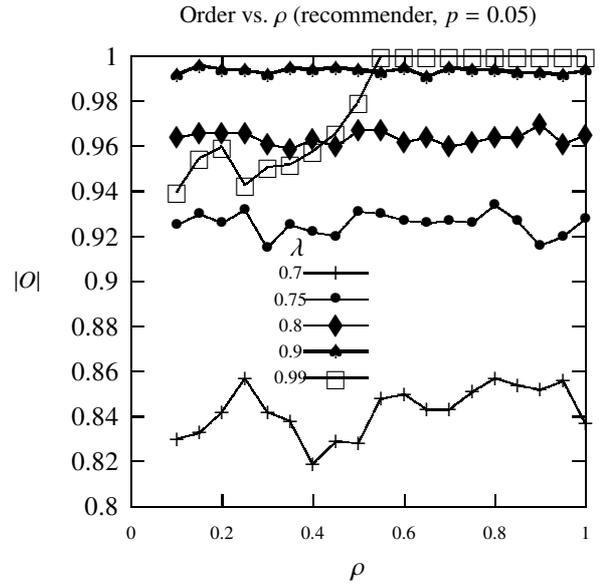

  \setlength{\unitlength}{0.240900pt}
  \ifx\plotpoint\undefined\newsavebox{\plotpoint}\fi
  \sbox{\plotpoint}{\rule[-0.200pt]{0.400pt}{0.400pt}}%

  \caption{$|O_a|$ plotted against $\rho$ for a system with limited
    range interaction and recommender for different values of the
    conviction parameter $\lambda$ (in the range $[\lambda_c,1)$.}
    \label{lim_R_rho}
\end{figure}

Figure~\ref{lim_R} shows the value $|O_a|$ plotted against $\lambda$
for several values of the localization parameter $\rho$.
Note that the dependence on $\rho$ has basically disappeared, and the
system has reverted to its full connection behavior. This is confirmed
in figure~\ref{lim_R_rho}, in which $|O_a|$ is plotted as a function
of $\rho$.
\begin{figure}[htb]
  \setlength{\unitlength}{0.240900pt}
  \ifx\plotpoint\undefined\newsavebox{\plotpoint}\fi
  \sbox{\plotpoint}{\rule[-0.200pt]{0.400pt}{0.400pt}}%
  %
  \caption{Maximum cluster size as a function of the conviction
    parameter for various values of the localization parameter
    and $\Omega=0.99$ in a system with recommender.}
  \label{clst_R}
\end{figure}
The variations of $|O_a|$ with $\rho$ are extremely limited (the $y$
axis of the figure covers only the range $[0.8,1]$, which makes the
irregularities seem stronger then they actually are). Even at this low
level of interaction with the recommender (5\%), and even though the
recommender allows interaction only of individuals with similar
opinions, it appears that the long range interactions are sufficient
to impede de formation of \dqt{pockets} of minority opinion. 

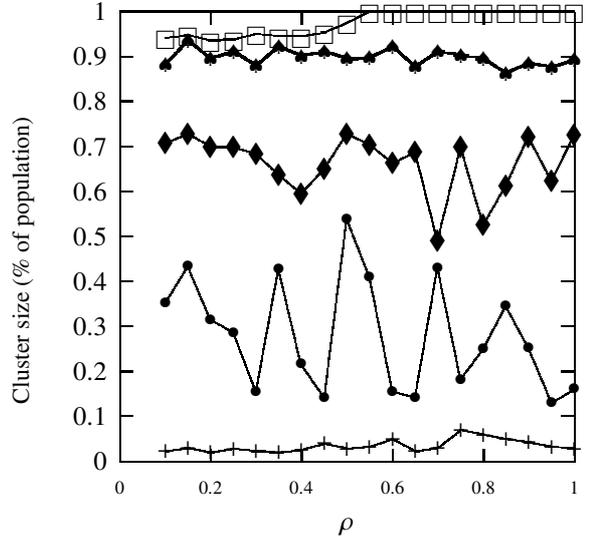
\begin{figure}[bht]
  \setlength{\unitlength}{0.240900pt}
  \ifx\plotpoint\undefined\newsavebox{\plotpoint}\fi
  \sbox{\plotpoint}{\rule[-0.200pt]{0.400pt}{0.400pt}}%
  \begin{picture}(960,960)(0,0)
    \sbox{\plotpoint}{\rule[-0.200pt]{0.400pt}{0.400pt}}%
    \put(182.0,131.0){\rule[-0.200pt]{4.818pt}{0.400pt}}
    \put(162,131){\makebox(0,0)[r]{ 0}}
    \put(868.0,131.0){\rule[-0.200pt]{4.818pt}{0.400pt}}
    \put(182.0,202.0){\rule[-0.200pt]{4.818pt}{0.400pt}}
    \put(162,202){\makebox(0,0)[r]{ 0.1}}
    \put(868.0,202.0){\rule[-0.200pt]{4.818pt}{0.400pt}}
    \put(182.0,272.0){\rule[-0.200pt]{4.818pt}{0.400pt}}
    \put(162,272){\makebox(0,0)[r]{ 0.2}}
    \put(868.0,272.0){\rule[-0.200pt]{4.818pt}{0.400pt}}
    \put(182.0,343.0){\rule[-0.200pt]{4.818pt}{0.400pt}}
    \put(162,343){\makebox(0,0)[r]{ 0.3}}
    \put(868.0,343.0){\rule[-0.200pt]{4.818pt}{0.400pt}}
    \put(182.0,413.0){\rule[-0.200pt]{4.818pt}{0.400pt}}
    \put(162,413){\makebox(0,0)[r]{ 0.4}}
    \put(868.0,413.0){\rule[-0.200pt]{4.818pt}{0.400pt}}
    \put(182.0,484.0){\rule[-0.200pt]{4.818pt}{0.400pt}}
    \put(162,484){\makebox(0,0)[r]{ 0.5}}
    \put(868.0,484.0){\rule[-0.200pt]{4.818pt}{0.400pt}}
    \put(182.0,554.0){\rule[-0.200pt]{4.818pt}{0.400pt}}
    \put(162,554){\makebox(0,0)[r]{ 0.6}}
    \put(868.0,554.0){\rule[-0.200pt]{4.818pt}{0.400pt}}
    \put(182.0,624.0){\rule[-0.200pt]{4.818pt}{0.400pt}}
    \put(162,624){\makebox(0,0)[r]{ 0.7}}
    \put(868.0,624.0){\rule[-0.200pt]{4.818pt}{0.400pt}}
    \put(182.0,695.0){\rule[-0.200pt]{4.818pt}{0.400pt}}
    \put(162,695){\makebox(0,0)[r]{ 0.8}}
    \put(868.0,695.0){\rule[-0.200pt]{4.818pt}{0.400pt}}
    \put(182.0,765.0){\rule[-0.200pt]{4.818pt}{0.400pt}}
    \put(162,765){\makebox(0,0)[r]{ 0.9}}
    \put(868.0,765.0){\rule[-0.200pt]{4.818pt}{0.400pt}}
    \put(182.0,836.0){\rule[-0.200pt]{4.818pt}{0.400pt}}
    \put(162,836){\makebox(0,0)[r]{ 1}}
    \put(868.0,836.0){\rule[-0.200pt]{4.818pt}{0.400pt}}
    \put(182.0,131.0){\rule[-0.200pt]{0.400pt}{4.818pt}}
    \put(182,90){\makebox(0,0){\scriptsize 0}}
    \put(182.0,816.0){\rule[-0.200pt]{0.400pt}{4.818pt}}
    \put(323.0,131.0){\rule[-0.200pt]{0.400pt}{4.818pt}}
    \put(323,90){\makebox(0,0){\scriptsize 0.2}}
    \put(323.0,816.0){\rule[-0.200pt]{0.400pt}{4.818pt}}
    \put(464.0,131.0){\rule[-0.200pt]{0.400pt}{4.818pt}}
    \put(464,90){\makebox(0,0){\scriptsize 0.4}}
    \put(464.0,816.0){\rule[-0.200pt]{0.400pt}{4.818pt}}
    \put(606.0,131.0){\rule[-0.200pt]{0.400pt}{4.818pt}}
    \put(606,90){\makebox(0,0){\scriptsize 0.6}}
    \put(606.0,816.0){\rule[-0.200pt]{0.400pt}{4.818pt}}
    \put(747.0,131.0){\rule[-0.200pt]{0.400pt}{4.818pt}}
    \put(747,90){\makebox(0,0){\scriptsize 0.8}}
    \put(747.0,816.0){\rule[-0.200pt]{0.400pt}{4.818pt}}
    \put(888.0,131.0){\rule[-0.200pt]{0.400pt}{4.818pt}}
    \put(888,90){\makebox(0,0){\scriptsize 1}}
    \put(888.0,816.0){\rule[-0.200pt]{0.400pt}{4.818pt}}
    \put(182.0,131.0){\rule[-0.200pt]{0.400pt}{169.834pt}}
    \put(182.0,131.0){\rule[-0.200pt]{170.075pt}{0.400pt}}
    \put(888.0,131.0){\rule[-0.200pt]{0.400pt}{169.834pt}}
    \put(182.0,836.0){\rule[-0.200pt]{170.075pt}{0.400pt}}
    \put(41,483){\makebox(0,0){\small{\begin{rotate}{90}\hspace{-7em}Cluster size (\% of population)\end{rotate}}}}
    \put(535,29){\makebox(0,0){$\rho$}}
    \put(535,898){\makebox(0,0){\small{Maximum cluster vs. $\rho$ ($\Omega=0.99$), recommender, $p=0.05$}}}
    \put(253,147){\usebox{\plotpoint}}
    \multiput(253.00,147.59)(3.827,0.477){7}{\rule{2.900pt}{0.115pt}}
    \multiput(253.00,146.17)(28.981,5.000){2}{\rule{1.450pt}{0.400pt}}
    \multiput(288.00,150.93)(2.628,-0.485){11}{\rule{2.100pt}{0.117pt}}
    \multiput(288.00,151.17)(30.641,-7.000){2}{\rule{1.050pt}{0.400pt}}
    \multiput(323.00,145.59)(3.203,0.482){9}{\rule{2.500pt}{0.116pt}}
    \multiput(323.00,144.17)(30.811,6.000){2}{\rule{1.250pt}{0.400pt}}
    \multiput(359.00,149.94)(5.014,-0.468){5}{\rule{3.600pt}{0.113pt}}
    \multiput(359.00,150.17)(27.528,-4.000){2}{\rule{1.800pt}{0.400pt}}
    \put(394,145.17){\rule{7.100pt}{0.400pt}}
    \multiput(394.00,146.17)(20.264,-2.000){2}{\rule{3.550pt}{0.400pt}}
    \multiput(429.00,145.60)(5.014,0.468){5}{\rule{3.600pt}{0.113pt}}
    \multiput(429.00,144.17)(27.528,4.000){2}{\rule{1.800pt}{0.400pt}}
    \multiput(464.00,149.58)(1.850,0.491){17}{\rule{1.540pt}{0.118pt}}
    \multiput(464.00,148.17)(32.804,10.000){2}{\rule{0.770pt}{0.400pt}}
    \multiput(500.00,157.93)(2.277,-0.488){13}{\rule{1.850pt}{0.117pt}}
    \multiput(500.00,158.17)(31.160,-8.000){2}{\rule{0.925pt}{0.400pt}}
    \multiput(535.00,151.61)(7.607,0.447){3}{\rule{4.767pt}{0.108pt}}
    \multiput(535.00,150.17)(25.107,3.000){2}{\rule{2.383pt}{0.400pt}}
    \multiput(570.00,154.58)(1.530,0.492){21}{\rule{1.300pt}{0.119pt}}
    \multiput(570.00,153.17)(33.302,12.000){2}{\rule{0.650pt}{0.400pt}}
    \multiput(606.00,164.92)(0.927,-0.495){35}{\rule{0.837pt}{0.119pt}}
    \multiput(606.00,165.17)(33.263,-19.000){2}{\rule{0.418pt}{0.400pt}}
    \multiput(641.00,147.59)(3.827,0.477){7}{\rule{2.900pt}{0.115pt}}
    \multiput(641.00,146.17)(28.981,5.000){2}{\rule{1.450pt}{0.400pt}}
    \multiput(676.00,152.58)(0.643,0.497){53}{\rule{0.614pt}{0.120pt}}
    \multiput(676.00,151.17)(34.725,28.000){2}{\rule{0.307pt}{0.400pt}}
    \multiput(712.00,178.93)(2.628,-0.485){11}{\rule{2.100pt}{0.117pt}}
    \multiput(712.00,179.17)(30.641,-7.000){2}{\rule{1.050pt}{0.400pt}}
    \multiput(747.00,171.93)(2.628,-0.485){11}{\rule{2.100pt}{0.117pt}}
    \multiput(747.00,172.17)(30.641,-7.000){2}{\rule{1.050pt}{0.400pt}}
    \multiput(782.00,164.93)(3.827,-0.477){7}{\rule{2.900pt}{0.115pt}}
    \multiput(782.00,165.17)(28.981,-5.000){2}{\rule{1.450pt}{0.400pt}}
    \multiput(817.00,159.93)(2.705,-0.485){11}{\rule{2.157pt}{0.117pt}}
    \multiput(817.00,160.17)(31.523,-7.000){2}{\rule{1.079pt}{0.400pt}}
    \multiput(853.00,152.95)(7.607,-0.447){3}{\rule{4.767pt}{0.108pt}}
    \multiput(853.00,153.17)(25.107,-3.000){2}{\rule{2.383pt}{0.400pt}}
    \put(253,147){\makebox(0,0){$+$}}
    \put(288,152){\makebox(0,0){$+$}}
    \put(323,145){\makebox(0,0){$+$}}
    \put(359,151){\makebox(0,0){$+$}}
    \put(394,147){\makebox(0,0){$+$}}
    \put(429,145){\makebox(0,0){$+$}}
    \put(464,149){\makebox(0,0){$+$}}
    \put(500,159){\makebox(0,0){$+$}}
    \put(535,151){\makebox(0,0){$+$}}
    \put(570,154){\makebox(0,0){$+$}}
    \put(606,166){\makebox(0,0){$+$}}
    \put(641,147){\makebox(0,0){$+$}}
    \put(676,152){\makebox(0,0){$+$}}
    \put(712,180){\makebox(0,0){$+$}}
    \put(747,173){\makebox(0,0){$+$}}
    \put(782,166){\makebox(0,0){$+$}}
    \put(817,161){\makebox(0,0){$+$}}
    \put(853,154){\makebox(0,0){$+$}}
    \put(888,151){\makebox(0,0){$+$}}
    \put(253,379){\usebox{\plotpoint}}
    \multiput(253.58,379.00)(0.498,0.816){67}{\rule{0.120pt}{0.751pt}}
    \multiput(252.17,379.00)(35.000,55.440){2}{\rule{0.400pt}{0.376pt}}
    \multiput(288.58,431.65)(0.498,-1.192){67}{\rule{0.120pt}{1.049pt}}
    \multiput(287.17,433.82)(35.000,-80.824){2}{\rule{0.400pt}{0.524pt}}
    \multiput(323.00,351.92)(0.862,-0.496){39}{\rule{0.786pt}{0.119pt}}
    \multiput(323.00,352.17)(34.369,-21.000){2}{\rule{0.393pt}{0.400pt}}
    \multiput(359.58,327.22)(0.498,-1.322){67}{\rule{0.120pt}{1.151pt}}
    \multiput(358.17,329.61)(35.000,-89.610){2}{\rule{0.400pt}{0.576pt}}
    \multiput(394.58,240.00)(0.498,2.765){67}{\rule{0.120pt}{2.294pt}}
    \multiput(393.17,240.00)(35.000,187.238){2}{\rule{0.400pt}{1.147pt}}
    \multiput(429.58,424.52)(0.498,-2.144){67}{\rule{0.120pt}{1.803pt}}
    \multiput(428.17,428.26)(35.000,-145.258){2}{\rule{0.400pt}{0.901pt}}
    \multiput(464.58,280.19)(0.498,-0.723){69}{\rule{0.120pt}{0.678pt}}
    \multiput(463.17,281.59)(36.000,-50.593){2}{\rule{0.400pt}{0.339pt}}
    \multiput(500.58,231.00)(0.498,4.020){67}{\rule{0.120pt}{3.289pt}}
    \multiput(499.17,231.00)(35.000,272.174){2}{\rule{0.400pt}{1.644pt}}
    \multiput(535.58,505.32)(0.498,-1.293){67}{\rule{0.120pt}{1.129pt}}
    \multiput(534.17,507.66)(35.000,-87.658){2}{\rule{0.400pt}{0.564pt}}
    \multiput(570.58,411.28)(0.498,-2.519){69}{\rule{0.120pt}{2.100pt}}
    \multiput(569.17,415.64)(36.000,-175.641){2}{\rule{0.400pt}{1.050pt}}
    \multiput(606.00,238.93)(2.009,-0.489){15}{\rule{1.656pt}{0.118pt}}
    \multiput(606.00,239.17)(31.564,-9.000){2}{\rule{0.828pt}{0.400pt}}
    \multiput(641.58,231.00)(0.498,2.924){67}{\rule{0.120pt}{2.420pt}}
    \multiput(640.17,231.00)(35.000,197.977){2}{\rule{0.400pt}{1.210pt}}
    \multiput(676.58,425.51)(0.498,-2.449){69}{\rule{0.120pt}{2.044pt}}
    \multiput(675.17,429.76)(36.000,-170.757){2}{\rule{0.400pt}{1.022pt}}
    \multiput(712.58,259.00)(0.498,0.686){67}{\rule{0.120pt}{0.649pt}}
    \multiput(711.17,259.00)(35.000,46.654){2}{\rule{0.400pt}{0.324pt}}
    \multiput(747.58,307.00)(0.498,0.961){67}{\rule{0.120pt}{0.866pt}}
    \multiput(746.17,307.00)(35.000,65.203){2}{\rule{0.400pt}{0.433pt}}
    \multiput(782.58,370.50)(0.498,-0.932){67}{\rule{0.120pt}{0.843pt}}
    \multiput(781.17,372.25)(35.000,-63.251){2}{\rule{0.400pt}{0.421pt}}
    \multiput(817.58,304.62)(0.498,-1.200){69}{\rule{0.120pt}{1.056pt}}
    \multiput(816.17,306.81)(36.000,-83.809){2}{\rule{0.400pt}{0.528pt}}
    \multiput(853.00,223.58)(0.837,0.496){39}{\rule{0.767pt}{0.119pt}}
    \multiput(853.00,222.17)(33.409,21.000){2}{\rule{0.383pt}{0.400pt}}
    \put(253,379){\makebox(0,0){$\bullet$}}
    \put(288,436){\makebox(0,0){$\bullet$}}
    \put(323,353){\makebox(0,0){$\bullet$}}
    \put(359,332){\makebox(0,0){$\bullet$}}
    \put(394,240){\makebox(0,0){$\bullet$}}
    \put(429,432){\makebox(0,0){$\bullet$}}
    \put(464,283){\makebox(0,0){$\bullet$}}
    \put(500,231){\makebox(0,0){$\bullet$}}
    \put(535,510){\makebox(0,0){$\bullet$}}
    \put(570,420){\makebox(0,0){$\bullet$}}
    \put(606,240){\makebox(0,0){$\bullet$}}
    \put(641,231){\makebox(0,0){$\bullet$}}
    \put(676,434){\makebox(0,0){$\bullet$}}
    \put(712,259){\makebox(0,0){$\bullet$}}
    \put(747,307){\makebox(0,0){$\bullet$}}
    \put(782,374){\makebox(0,0){$\bullet$}}
    \put(817,309){\makebox(0,0){$\bullet$}}
    \put(853,223){\makebox(0,0){$\bullet$}}
    \put(888,244){\makebox(0,0){$\bullet$}}
    \put(253,630){\usebox{\plotpoint}}
    \multiput(253.00,630.58)(1.268,0.494){25}{\rule{1.100pt}{0.119pt}}
    \multiput(253.00,629.17)(32.717,14.000){2}{\rule{0.550pt}{0.400pt}}
    \multiput(288.00,642.92)(0.880,-0.496){37}{\rule{0.800pt}{0.119pt}}
    \multiput(288.00,643.17)(33.340,-20.000){2}{\rule{0.400pt}{0.400pt}}
    \put(323,622.67){\rule{8.672pt}{0.400pt}}
    \multiput(323.00,623.17)(18.000,-1.000){2}{\rule{4.336pt}{0.400pt}}
    \multiput(359.00,621.92)(1.628,-0.492){19}{\rule{1.373pt}{0.118pt}}
    \multiput(359.00,622.17)(32.151,-11.000){2}{\rule{0.686pt}{0.400pt}}
    \multiput(394.00,610.92)(0.546,-0.497){61}{\rule{0.538pt}{0.120pt}}
    \multiput(394.00,611.17)(33.884,-32.000){2}{\rule{0.269pt}{0.400pt}}
    \multiput(429.00,578.92)(0.583,-0.497){57}{\rule{0.567pt}{0.120pt}}
    \multiput(429.00,579.17)(33.824,-30.000){2}{\rule{0.283pt}{0.400pt}}
    \multiput(464.58,550.00)(0.498,0.541){69}{\rule{0.120pt}{0.533pt}}
    \multiput(463.17,550.00)(36.000,37.893){2}{\rule{0.400pt}{0.267pt}}
    \multiput(500.58,589.00)(0.498,0.788){67}{\rule{0.120pt}{0.729pt}}
    \multiput(499.17,589.00)(35.000,53.488){2}{\rule{0.400pt}{0.364pt}}
    \multiput(535.00,642.92)(1.039,-0.495){31}{\rule{0.924pt}{0.119pt}}
    \multiput(535.00,643.17)(33.083,-17.000){2}{\rule{0.462pt}{0.400pt}}
    \multiput(570.00,625.92)(0.621,-0.497){55}{\rule{0.597pt}{0.120pt}}
    \multiput(570.00,626.17)(34.762,-29.000){2}{\rule{0.298pt}{0.400pt}}
    \multiput(606.00,598.58)(0.980,0.495){33}{\rule{0.878pt}{0.119pt}}
    \multiput(606.00,597.17)(33.178,18.000){2}{\rule{0.439pt}{0.400pt}}
    \multiput(641.58,608.94)(0.498,-2.014){67}{\rule{0.120pt}{1.700pt}}
    \multiput(640.17,612.47)(35.000,-136.472){2}{\rule{0.400pt}{0.850pt}}
    \multiput(676.58,476.00)(0.498,2.056){69}{\rule{0.120pt}{1.733pt}}
    \multiput(675.17,476.00)(36.000,143.402){2}{\rule{0.400pt}{0.867pt}}
    \multiput(712.58,616.80)(0.498,-1.755){67}{\rule{0.120pt}{1.494pt}}
    \multiput(711.17,619.90)(35.000,-118.899){2}{\rule{0.400pt}{0.747pt}}
    \multiput(747.58,501.00)(0.498,0.889){67}{\rule{0.120pt}{0.809pt}}
    \multiput(746.17,501.00)(35.000,60.322){2}{\rule{0.400pt}{0.404pt}}
    \multiput(782.58,563.00)(0.498,1.091){67}{\rule{0.120pt}{0.969pt}}
    \multiput(781.17,563.00)(35.000,73.990){2}{\rule{0.400pt}{0.484pt}}
    \multiput(817.58,635.40)(0.498,-0.962){69}{\rule{0.120pt}{0.867pt}}
    \multiput(816.17,637.20)(36.000,-67.201){2}{\rule{0.400pt}{0.433pt}}
    \multiput(853.58,570.00)(0.498,1.033){67}{\rule{0.120pt}{0.923pt}}
    \multiput(852.17,570.00)(35.000,70.085){2}{\rule{0.400pt}{0.461pt}}
    \put(253,630){\makebox(0,0){$\blacklozenge$}}
    \put(288,644){\makebox(0,0){$\blacklozenge$}}
    \put(323,624){\makebox(0,0){$\blacklozenge$}}
    \put(359,623){\makebox(0,0){$\blacklozenge$}}
    \put(394,612){\makebox(0,0){$\blacklozenge$}}
    \put(429,580){\makebox(0,0){$\blacklozenge$}}
    \put(464,550){\makebox(0,0){$\blacklozenge$}}
    \put(500,589){\makebox(0,0){$\blacklozenge$}}
    \put(535,644){\makebox(0,0){$\blacklozenge$}}
    \put(570,627){\makebox(0,0){$\blacklozenge$}}
    \put(606,598){\makebox(0,0){$\blacklozenge$}}
    \put(641,616){\makebox(0,0){$\blacklozenge$}}
    \put(676,476){\makebox(0,0){$\blacklozenge$}}
    \put(712,623){\makebox(0,0){$\blacklozenge$}}
    \put(747,501){\makebox(0,0){$\blacklozenge$}}
    \put(782,563){\makebox(0,0){$\blacklozenge$}}
    \put(817,639){\makebox(0,0){$\blacklozenge$}}
    \put(853,570){\makebox(0,0){$\blacklozenge$}}
    \put(888,642){\makebox(0,0){$\blacklozenge$}}
    \sbox{\plotpoint}{\rule[-0.400pt]{0.800pt}{0.800pt}}%
    \put(253,753){\usebox{\plotpoint}}
    \multiput(254.41,753.00)(0.503,0.527){63}{\rule{0.121pt}{1.046pt}}
    \multiput(251.34,753.00)(35.000,34.830){2}{\rule{0.800pt}{0.523pt}}
    \multiput(288.00,788.09)(0.625,-0.504){49}{\rule{1.200pt}{0.121pt}}
    \multiput(288.00,788.34)(32.509,-28.000){2}{\rule{0.600pt}{0.800pt}}
    \multiput(323.00,763.40)(1.736,0.512){15}{\rule{2.818pt}{0.123pt}}
    \multiput(323.00,760.34)(30.151,11.000){2}{\rule{1.409pt}{0.800pt}}
    \multiput(359.00,771.09)(0.802,-0.505){37}{\rule{1.473pt}{0.122pt}}
    \multiput(359.00,771.34)(31.943,-22.000){2}{\rule{0.736pt}{0.800pt}}
    \multiput(394.00,752.41)(0.583,0.503){53}{\rule{1.133pt}{0.121pt}}
    \multiput(394.00,749.34)(32.648,30.000){2}{\rule{0.567pt}{0.800pt}}
    \multiput(429.00,779.09)(1.201,-0.508){23}{\rule{2.067pt}{0.122pt}}
    \multiput(429.00,779.34)(30.711,-15.000){2}{\rule{1.033pt}{0.800pt}}
    \multiput(464.00,767.40)(3.015,0.526){7}{\rule{4.314pt}{0.127pt}}
    \multiput(464.00,764.34)(27.045,7.000){2}{\rule{2.157pt}{0.800pt}}
    \multiput(500.00,771.08)(1.686,-0.512){15}{\rule{2.745pt}{0.123pt}}
    \multiput(500.00,771.34)(29.302,-11.000){2}{\rule{1.373pt}{0.800pt}}
    \put(535,760.84){\rule{8.432pt}{0.800pt}}
    \multiput(535.00,760.34)(17.500,1.000){2}{\rule{4.216pt}{0.800pt}}
    \multiput(570.00,764.41)(1.019,0.506){29}{\rule{1.800pt}{0.122pt}}
    \multiput(570.00,761.34)(32.264,18.000){2}{\rule{0.900pt}{0.800pt}}
    \multiput(606.00,779.09)(0.545,-0.503){57}{\rule{1.075pt}{0.121pt}}
    \multiput(606.00,779.34)(32.769,-32.000){2}{\rule{0.538pt}{0.800pt}}
    \multiput(641.00,750.41)(0.733,0.504){41}{\rule{1.367pt}{0.122pt}}
    \multiput(641.00,747.34)(32.163,24.000){2}{\rule{0.683pt}{0.800pt}}
    \multiput(676.00,771.07)(3.811,-0.536){5}{\rule{5.000pt}{0.129pt}}
    \multiput(676.00,771.34)(25.622,-6.000){2}{\rule{2.500pt}{0.800pt}}
    \multiput(712.00,765.06)(5.462,-0.560){3}{\rule{5.800pt}{0.135pt}}
    \multiput(712.00,765.34)(22.962,-5.000){2}{\rule{2.900pt}{0.800pt}}
    \multiput(747.00,760.09)(0.766,-0.505){39}{\rule{1.417pt}{0.122pt}}
    \multiput(747.00,760.34)(32.058,-23.000){2}{\rule{0.709pt}{0.800pt}}
    \multiput(782.00,740.41)(1.121,0.507){25}{\rule{1.950pt}{0.122pt}}
    \multiput(782.00,737.34)(30.953,16.000){2}{\rule{0.975pt}{0.800pt}}
    \multiput(817.00,753.08)(3.015,-0.526){7}{\rule{4.314pt}{0.127pt}}
    \multiput(817.00,753.34)(27.045,-7.000){2}{\rule{2.157pt}{0.800pt}}
    \multiput(853.00,749.41)(1.530,0.511){17}{\rule{2.533pt}{0.123pt}}
    \multiput(853.00,746.34)(29.742,12.000){2}{\rule{1.267pt}{0.800pt}}
    \put(253,753){\makebox(0,0){$\spadesuit$}}
    \put(288,790){\makebox(0,0){$\spadesuit$}}
    \put(323,762){\makebox(0,0){$\spadesuit$}}
    \put(359,773){\makebox(0,0){$\spadesuit$}}
    \put(394,751){\makebox(0,0){$\spadesuit$}}
    \put(429,781){\makebox(0,0){$\spadesuit$}}
    \put(464,766){\makebox(0,0){$\spadesuit$}}
    \put(500,773){\makebox(0,0){$\spadesuit$}}
    \put(535,762){\makebox(0,0){$\spadesuit$}}
    \put(570,763){\makebox(0,0){$\spadesuit$}}
    \put(606,781){\makebox(0,0){$\spadesuit$}}
    \put(641,749){\makebox(0,0){$\spadesuit$}}
    \put(676,773){\makebox(0,0){$\spadesuit$}}
    \put(712,767){\makebox(0,0){$\spadesuit$}}
    \put(747,762){\makebox(0,0){$\spadesuit$}}
    \put(782,739){\makebox(0,0){$\spadesuit$}}
    \put(817,755){\makebox(0,0){$\spadesuit$}}
    \put(853,748){\makebox(0,0){$\spadesuit$}}
    \put(888,760){\makebox(0,0){$\spadesuit$}}
    \sbox{\plotpoint}{\rule[-0.200pt]{0.400pt}{0.400pt}}%
    \put(253,794){\usebox{\plotpoint}}
    \multiput(253.00,794.59)(3.827,0.477){7}{\rule{2.900pt}{0.115pt}}
    \multiput(253.00,793.17)(28.981,5.000){2}{\rule{1.450pt}{0.400pt}}
    \multiput(288.00,797.93)(2.009,-0.489){15}{\rule{1.656pt}{0.118pt}}
    \multiput(288.00,798.17)(31.564,-9.000){2}{\rule{0.828pt}{0.400pt}}
    \put(323,790.17){\rule{7.300pt}{0.400pt}}
    \multiput(323.00,789.17)(20.848,2.000){2}{\rule{3.650pt}{0.400pt}}
    \multiput(359.00,792.59)(2.009,0.489){15}{\rule{1.656pt}{0.118pt}}
    \multiput(359.00,791.17)(31.564,9.000){2}{\rule{0.828pt}{0.400pt}}
    \multiput(394.00,799.94)(5.014,-0.468){5}{\rule{3.600pt}{0.113pt}}
    \multiput(394.00,800.17)(27.528,-4.000){2}{\rule{1.800pt}{0.400pt}}
    \multiput(464.00,797.59)(3.203,0.482){9}{\rule{2.500pt}{0.116pt}}
    \multiput(464.00,796.17)(30.811,6.000){2}{\rule{1.250pt}{0.400pt}}
    \multiput(500.00,803.58)(1.181,0.494){27}{\rule{1.033pt}{0.119pt}}
    \multiput(500.00,802.17)(32.855,15.000){2}{\rule{0.517pt}{0.400pt}}
    \multiput(535.00,818.58)(0.980,0.495){33}{\rule{0.878pt}{0.119pt}}
    \multiput(535.00,817.17)(33.178,18.000){2}{\rule{0.439pt}{0.400pt}}
    \put(429.0,797.0){\rule[-0.200pt]{8.431pt}{0.400pt}}
    \put(253,794){\raisebox{-.8pt}{\makebox(0,0){$\Box$}}}
    \put(288,799){\raisebox{-.8pt}{\makebox(0,0){$\Box$}}}
    \put(323,790){\raisebox{-.8pt}{\makebox(0,0){$\Box$}}}
    \put(359,792){\raisebox{-.8pt}{\makebox(0,0){$\Box$}}}
    \put(394,801){\raisebox{-.8pt}{\makebox(0,0){$\Box$}}}
    \put(429,797){\raisebox{-.8pt}{\makebox(0,0){$\Box$}}}
    \put(464,797){\raisebox{-.8pt}{\makebox(0,0){$\Box$}}}
    \put(500,803){\raisebox{-.8pt}{\makebox(0,0){$\Box$}}}
    \put(535,818){\raisebox{-.8pt}{\makebox(0,0){$\Box$}}}
    \put(570,836){\raisebox{-.8pt}{\makebox(0,0){$\Box$}}}
    \put(606,836){\raisebox{-.8pt}{\makebox(0,0){$\Box$}}}
    \put(641,836){\raisebox{-.8pt}{\makebox(0,0){$\Box$}}}
    \put(676,836){\raisebox{-.8pt}{\makebox(0,0){$\Box$}}}
    \put(712,836){\raisebox{-.8pt}{\makebox(0,0){$\Box$}}}
    \put(747,836){\raisebox{-.8pt}{\makebox(0,0){$\Box$}}}
    \put(782,836){\raisebox{-.8pt}{\makebox(0,0){$\Box$}}}
    \put(817,836){\raisebox{-.8pt}{\makebox(0,0){$\Box$}}}
    \put(853,836){\raisebox{-.8pt}{\makebox(0,0){$\Box$}}}
    \put(888,836){\raisebox{-.8pt}{\makebox(0,0){$\Box$}}}
    \put(570.0,836.0){\rule[-0.200pt]{76.606pt}{0.400pt}}
    \put(182.0,131.0){\rule[-0.200pt]{0.400pt}{169.834pt}}
    \put(182.0,131.0){\rule[-0.200pt]{170.075pt}{0.400pt}}
    \put(888.0,131.0){\rule[-0.200pt]{0.400pt}{169.834pt}}
    \put(182.0,836.0){\rule[-0.200pt]{170.075pt}{0.400pt}}
          \end{picture}
  \caption{Maximum cluster size as a function of the localization
    parameter $\rho$ for various values of the conviction parameter
    $\lambda$ and $\Omega=0.99$ in a system with recommender. The
    value of $\lambda$ for the five curves, from top to bottom, are
    $0.99$, $0.9$, $0.8$, $0.75$, $0.7$.}
  \label{clst_R_rho}
\end{figure}

This is confirmed if we analyze the size of the largest cluster in the
system (figures~\ref{clst_R},\ref{clst_R_rho}). Figure~\ref{clst_R}
shows a percolation effect for $\lambda\approx{0.7}$ analogous to that
of figure~\ref{clst_O} and virtually independent of the localization
parameter $\rho$.
The effects of the recommendation system are very evident if we
compare figure~\ref{clst_R_rho} (maximum cluster size as a function of
the localization parameter) with figure~\ref{clst_rho}. In the absence
of recommender, there is, for $\lambda>\lambda_c$, a sharp transition
in the cluster size corresponding to $\rho\approx{0.5}$.
With the reccomender system, no such transition is present. There is,
as in the case of absence of recommender, an instability zone
corresponding to $\lambda\approx{0.75}$ (right in the middle of the
symmetry breaking transition of the conviction parameter), but the
overall behavior is quite independent of $\rho$. 

The results of the model indicate that the presence of a recommender
system, even one used in only 5\% of the interaction is cabable of
inducing a global consensus in a comunity with only local
interactions, leading to a symmetry breaking transition of the order
parameter and to a percolation effect of the opinion clusters similar
to those characteristics of a fully connected system. 

We must remark that a \emph{fair} recommender system like the one used
so far induces a global opinion but it doesn't constraint the nature
of this consensus. In the community with recommender, just like in the
ones without it, after the symmetry break ($\lambda\approx{1}$) the
values $O_a=1$ and $O_a=-1$ occurr with probability $1/2$.

\subsection{Mischievous Recommenders}
In order to determine in what condition the breaking of symmetry may
be accompanied by a polarization of opinions, we create a system with
a \emph{mischievous} recommender system, that is, a system that tries to
polarize the community towards the opinion $O_a=1$. In order to do
this, given an individual $i$ that is interacting with the
recommender, the recommender determines the value $r(i)$ as in
(\ref{recom1}). If $O_{r(i)}(t)>O_i(t)$, then the recommender uses
$O_{r(i)}$ as in the fair recommended case. If $O_{r(i)}(t)<O_i(t)$,
then the recommender \dqt{flips} $O_{r(i)}$ around $O_i$, replacing it
with $2O_i-O_{r(i)}$. This means that individual $i$ will always
receive at time $t$ a recomendation with a value greater than $O(t)$.

\begin{figure}[bht]
  \setlength{\unitlength}{0.240900pt}
  \ifx\plotpoint\undefined\newsavebox{\plotpoint}\fi
  \sbox{\plotpoint}{\rule[-0.200pt]{0.400pt}{0.400pt}}%
  \begin{picture}(960,960)(0,0)
    \sbox{\plotpoint}{\rule[-0.200pt]{0.400pt}{0.400pt}}%
    \put(182.0,131.0){\rule[-0.200pt]{4.818pt}{0.400pt}}
    \put(162,131){\makebox(0,0)[r]{ 0}}
    \put(868.0,131.0){\rule[-0.200pt]{4.818pt}{0.400pt}}
    \put(182.0,259.0){\rule[-0.200pt]{4.818pt}{0.400pt}}
    \put(162,259){\makebox(0,0)[r]{ 0.2}}
    \put(868.0,259.0){\rule[-0.200pt]{4.818pt}{0.400pt}}
    \put(182.0,387.0){\rule[-0.200pt]{4.818pt}{0.400pt}}
    \put(162,387){\makebox(0,0)[r]{ 0.4}}
    \put(868.0,387.0){\rule[-0.200pt]{4.818pt}{0.400pt}}
    \put(182.0,516.0){\rule[-0.200pt]{4.818pt}{0.400pt}}
    \put(162,516){\makebox(0,0)[r]{ 0.6}}
    \put(868.0,516.0){\rule[-0.200pt]{4.818pt}{0.400pt}}
    \put(182.0,644.0){\rule[-0.200pt]{4.818pt}{0.400pt}}
    \put(162,644){\makebox(0,0)[r]{ 0.8}}
    \put(868.0,644.0){\rule[-0.200pt]{4.818pt}{0.400pt}}
    \put(182.0,772.0){\rule[-0.200pt]{4.818pt}{0.400pt}}
    \put(162,772){\makebox(0,0)[r]{ 1}}
    \put(868.0,772.0){\rule[-0.200pt]{4.818pt}{0.400pt}}
    \put(182.0,131.0){\rule[-0.200pt]{0.400pt}{4.818pt}}
    \put(182,90){\makebox(0,0){\scriptsize 0}}
    \put(182.0,816.0){\rule[-0.200pt]{0.400pt}{4.818pt}}
    \put(323.0,131.0){\rule[-0.200pt]{0.400pt}{4.818pt}}
    \put(323,90){\makebox(0,0){\scriptsize 0.2}}
    \put(323.0,816.0){\rule[-0.200pt]{0.400pt}{4.818pt}}
    \put(464.0,131.0){\rule[-0.200pt]{0.400pt}{4.818pt}}
    \put(464,90){\makebox(0,0){\scriptsize 0.4}}
    \put(464.0,816.0){\rule[-0.200pt]{0.400pt}{4.818pt}}
    \put(606.0,131.0){\rule[-0.200pt]{0.400pt}{4.818pt}}
    \put(606,90){\makebox(0,0){\scriptsize 0.6}}
    \put(606.0,816.0){\rule[-0.200pt]{0.400pt}{4.818pt}}
    \put(747.0,131.0){\rule[-0.200pt]{0.400pt}{4.818pt}}
    \put(747,90){\makebox(0,0){\scriptsize 0.8}}
    \put(747.0,816.0){\rule[-0.200pt]{0.400pt}{4.818pt}}
    \put(888.0,131.0){\rule[-0.200pt]{0.400pt}{4.818pt}}
    \put(888,90){\makebox(0,0){\scriptsize 1}}
    \put(888.0,816.0){\rule[-0.200pt]{0.400pt}{4.818pt}}
    \put(182.0,131.0){\rule[-0.200pt]{0.400pt}{169.834pt}}
    \put(182.0,131.0){\rule[-0.200pt]{170.075pt}{0.400pt}}
    \put(888.0,131.0){\rule[-0.200pt]{0.400pt}{169.834pt}}
    \put(182.0,836.0){\rule[-0.200pt]{170.075pt}{0.400pt}}
    \put(41,483){\makebox(0,0){\small{$P[O_a>0]$}}}
    \put(535,29){\makebox(0,0){$\lambda$}}
    \put(535,898){\makebox(0,0){\small{$P[O_a>0]$ vs.\ $\lambda$ (recommender)}}}
    \put(254,366){\makebox(0,0){$\rho$}}
    \put(254,326){\makebox(0,0){\scriptsize 0.05}}
    \put(274.0,326.0){\rule[-0.200pt]{24.090pt}{0.400pt}}
    \put(323,772){\usebox{\plotpoint}}
    \multiput(641.58,756.21)(0.498,-4.670){67}{\rule{0.120pt}{3.803pt}}
    \multiput(640.17,764.11)(35.000,-316.107){2}{\rule{0.400pt}{1.901pt}}
    \multiput(676.00,446.92)(1.305,-0.494){25}{\rule{1.129pt}{0.119pt}}
    \multiput(676.00,447.17)(33.658,-14.000){2}{\rule{0.564pt}{0.400pt}}
    \multiput(712.00,434.59)(2.628,0.485){11}{\rule{2.100pt}{0.117pt}}
    \multiput(712.00,433.17)(30.641,7.000){2}{\rule{1.050pt}{0.400pt}}
    \multiput(747.00,441.60)(5.014,0.468){5}{\rule{3.600pt}{0.113pt}}
    \multiput(747.00,440.17)(27.528,4.000){2}{\rule{1.800pt}{0.400pt}}
    \multiput(782.00,445.59)(3.112,0.482){9}{\rule{2.433pt}{0.116pt}}
    \multiput(782.00,444.17)(29.949,6.000){2}{\rule{1.217pt}{0.400pt}}
    \multiput(817.00,451.58)(1.215,0.494){27}{\rule{1.060pt}{0.119pt}}
    \multiput(817.00,450.17)(33.800,15.000){2}{\rule{0.530pt}{0.400pt}}
    \put(323,772){\makebox(0,0){$+$}}
    \put(464,772){\makebox(0,0){$+$}}
    \put(606,772){\makebox(0,0){$+$}}
    \put(641,772){\makebox(0,0){$+$}}
    \put(676,448){\makebox(0,0){$+$}}
    \put(712,434){\makebox(0,0){$+$}}
    \put(747,441){\makebox(0,0){$+$}}
    \put(782,445){\makebox(0,0){$+$}}
    \put(817,451){\makebox(0,0){$+$}}
    \put(853,466){\makebox(0,0){$+$}}
    \put(324,326){\makebox(0,0){$+$}}
    \put(323.0,772.0){\rule[-0.200pt]{76.606pt}{0.400pt}}
    \put(254,285){\makebox(0,0){\scriptsize 0.1}}
    \put(274.0,285.0){\rule[-0.200pt]{24.090pt}{0.400pt}}
    \put(323,772){\usebox{\plotpoint}}
    \multiput(641.58,756.97)(0.498,-4.439){67}{\rule{0.120pt}{3.620pt}}
    \multiput(640.17,764.49)(35.000,-300.487){2}{\rule{0.400pt}{1.810pt}}
    \multiput(676.00,464.59)(2.705,0.485){11}{\rule{2.157pt}{0.117pt}}
    \multiput(676.00,463.17)(31.523,7.000){2}{\rule{1.079pt}{0.400pt}}
    \multiput(712.00,469.92)(0.546,-0.497){61}{\rule{0.538pt}{0.120pt}}
    \multiput(712.00,470.17)(33.884,-32.000){2}{\rule{0.269pt}{0.400pt}}
    \multiput(747.00,439.61)(7.607,0.447){3}{\rule{4.767pt}{0.108pt}}
    \multiput(747.00,438.17)(25.107,3.000){2}{\rule{2.383pt}{0.400pt}}
    \multiput(782.00,442.58)(1.105,0.494){29}{\rule{0.975pt}{0.119pt}}
    \multiput(782.00,441.17)(32.976,16.000){2}{\rule{0.488pt}{0.400pt}}
    \multiput(817.00,456.92)(1.305,-0.494){25}{\rule{1.129pt}{0.119pt}}
    \multiput(817.00,457.17)(33.658,-14.000){2}{\rule{0.564pt}{0.400pt}}
    \put(323,772){\makebox(0,0){$\bullet$}}
    \put(464,772){\makebox(0,0){$\bullet$}}
    \put(606,772){\makebox(0,0){$\bullet$}}
    \put(641,772){\makebox(0,0){$\bullet$}}
    \put(676,464){\makebox(0,0){$\bullet$}}
    \put(712,471){\makebox(0,0){$\bullet$}}
    \put(747,439){\makebox(0,0){$\bullet$}}
    \put(782,442){\makebox(0,0){$\bullet$}}
    \put(817,458){\makebox(0,0){$\bullet$}}
    \put(853,444){\makebox(0,0){$\bullet$}}
    \put(324,285){\makebox(0,0){$\bullet$}}
    \put(323.0,772.0){\rule[-0.200pt]{76.606pt}{0.400pt}}
    \put(254,244){\makebox(0,0){\scriptsize 0.250}}
    \put(274.0,244.0){\rule[-0.200pt]{24.090pt}{0.400pt}}
    \put(323,772){\usebox{\plotpoint}}
    \multiput(641.58,756.02)(0.498,-4.728){67}{\rule{0.120pt}{3.849pt}}
    \multiput(640.17,764.01)(35.000,-320.012){2}{\rule{0.400pt}{1.924pt}}
    \multiput(676.00,444.58)(0.905,0.496){37}{\rule{0.820pt}{0.119pt}}
    \multiput(676.00,443.17)(34.298,20.000){2}{\rule{0.410pt}{0.400pt}}
    \multiput(712.00,464.58)(0.499,0.498){67}{\rule{0.500pt}{0.120pt}}
    \multiput(712.00,463.17)(33.962,35.000){2}{\rule{0.250pt}{0.400pt}}
    \multiput(747.58,496.73)(0.498,-0.557){67}{\rule{0.120pt}{0.546pt}}
    \multiput(746.17,497.87)(35.000,-37.867){2}{\rule{0.400pt}{0.273pt}}
    \multiput(782.00,458.92)(1.268,-0.494){25}{\rule{1.100pt}{0.119pt}}
    \multiput(782.00,459.17)(32.717,-14.000){2}{\rule{0.550pt}{0.400pt}}
    \multiput(817.00,446.58)(1.069,0.495){31}{\rule{0.947pt}{0.119pt}}
    \multiput(817.00,445.17)(34.034,17.000){2}{\rule{0.474pt}{0.400pt}}
    \put(323,772){\makebox(0,0){$\blacklozenge$}}
    \put(464,772){\makebox(0,0){$\blacklozenge$}}
    \put(606,772){\makebox(0,0){$\blacklozenge$}}
    \put(641,772){\makebox(0,0){$\blacklozenge$}}
    \put(676,444){\makebox(0,0){$\blacklozenge$}}
    \put(712,464){\makebox(0,0){$\blacklozenge$}}
    \put(747,499){\makebox(0,0){$\blacklozenge$}}
    \put(782,460){\makebox(0,0){$\blacklozenge$}}
    \put(817,446){\makebox(0,0){$\blacklozenge$}}
    \put(853,463){\makebox(0,0){$\blacklozenge$}}
    \put(324,244){\makebox(0,0){$\blacklozenge$}}
    \put(323.0,772.0){\rule[-0.200pt]{76.606pt}{0.400pt}}
    \put(182.0,131.0){\rule[-0.200pt]{0.400pt}{169.834pt}}
    \put(182.0,131.0){\rule[-0.200pt]{170.075pt}{0.400pt}}
    \put(888.0,131.0){\rule[-0.200pt]{0.400pt}{169.834pt}}
    \put(182.0,836.0){\rule[-0.200pt]{170.075pt}{0.400pt}}
  \end{picture}
  \caption{Probability of the symmetry breaking transition resulting in
    $O_a=1$ with a \dqt{mischievous} recommender as a function of the
    conviction parameter $\lambda$ for various values of the locality
    parameters $\rho$.}
  \label{unfair_bal}
\end{figure}
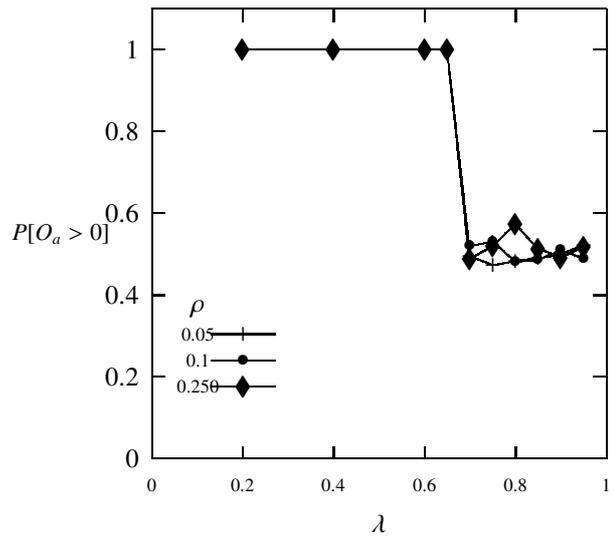

The results are synthetized in figure~\ref{unfair_bal}.
Whenever $\lambda<\lambda_c$, except for very low values of $\rho$,
with probability $1$, the outcome of the symmetry break is $O_a=1$,
that is, the mischievous recommender always succeeds in influencing the
dominant opinion of the community. As the onviction parameter moves
beyond the critical value $\lambda_c$, the probability that the
recommender may influence the community becomes smaller, until the
system reverts to the completely random behavior.

From these results it appears therefore that the presence of a
recommender system has always the effect of inducing a common
consensus, but that its capacity of directing this consensus is
limited to the disorganized area in which, in the absence of the
recommender, no opinion would prevail. 

\section{Conclusion}
In this paper we have studied a model of opinion formation based on
kinetic exchange. We have first shown that the symmetry breaking and
cluster percolation transitions, characteristics of the fully
connected system are lost if the interactions between individuals are
restricted to a neighborhood with a locality parameter
$\rho<{0.5}$. However, the presence of a recommender system is capable
of creating a system with the same characteristics of the fully
connected even with a probability of use of recommendation as little
as $0.01$ and with a locality as little as $0.1$.

Although the presence of the recommender does restore the symmetry
breaking transition and therefore creates a unique opinion in the
whole comunity, it does not, by its mere presence, determine the
nature of this opinion. A \emph{mischievous} system, one that would try
to direct the opinion to a particular state would only succeed in
pre-critical communities, that is, in communities in which the
conviction parameter $\lambda$ is less than the critical value
necessary for the symmetry breaking transition.


\end{document}